%% file: swifto2.tex
\documentclass[twocolumn,times]{aastex62}

\usepackage{float}
\usepackage{amsmath}
\usepackage{rotating}
\usepackage{amssymb}

\newcommand{\swift}{\textit{Swift}}

\newcommand{\xmm}{\textit{XMM-Newton}}

\newcommand{\fermi}{\textit{Fermi}}

\newcommand{\nustar}{\textit{NuSTAR}}

\newcommand{\indentitem}{\setlength\itemindent{25pt}}

\graphicspath{{./}{figures/}}

\received{}
\revised{}
\accepted{}
\submitjournal{ApJ}

\shorttitle{\swift-XRT GW Follow-up in O2}
\shortauthors{Klingler et al.}

\begin{document}

\title{{\em Swift}-XRT Follow-up of Gravitational Wave Triggers in the Second Advanced LIGO/Virgo Observing Run}

\correspondingauthor{N.~J. Klingler}
\email{njk5441@psu.edu}

\author[0000-0002-7465-0941]{N.~J.~Klingler}
\affiliation{Department of Astronomy and Astrophysics, The Pennsylvania State University, 525 Davey Laboratory, University Park, PA 16802, USA}
\author[0000-0002-6745-4790]{J.~A.~Kennea}
\affiliation{Department of Astronomy and Astrophysics, The Pennsylvania State University, 525 Davey Laboratory, University Park, PA 16802, USA}
\author[0000-0002-8465-3353]{P.~A.~Evans}
\affiliation{Department of Physics and Astronomy, University of Leicester, University Road, Leicester, LE1 7RH, UK}
\author[0000-0002-2810-8764]{A.~Tohuvavohu}
\affiliation{Department of Astronomy and Astrophysics, The Pennsylvania State University, 525 Davey Laboratory, University Park, PA 16802, USA}
\author[0000-0002-2810-8764]{S.~B.~Cenko}
\affiliation{Astrophysics Science Division, NASA Goddard Space Flight Center, Greenbelt MD, 20771 USA}
\affiliation{Joint Space-Science Institute, Computer and Space Sciences Building, University of Maryland, College Park, MD 20742, USA}
\author{S.~D.~Barthelmy}
\affiliation{Astrophysics Science Division, NASA Goddard Space Flight Center, Greenbelt MD, 20771 USA}
\author{A.~P.~Beardmore}
\affiliation{Department of Physics and Astronomy, University of Leicester, University Road, Leicester, LE1 7RH, UK}
\author{A.~A.~Breeveld}
\affiliation{University College London, Mullard Space Science Laboratory, Holmbury St. Mary, Dorking, RH5 6NT, U.K.}
\author[0000-0001-6272-5507]{P.~J.~Brown}
\affiliation{George P.\ and Cynthia Woods Mitchell Institute for Fundamental Physics \& Astronomy, Mitchell Physics Building, Texas A.~\& M.~University, Department of Physics and Astronomy, College Station, TX 77843, USA}
\author{D.~N.~Burrows}
\affiliation{Department of Astronomy and Astrophysics, The Pennsylvania State University, 525 Davey Laboratory, University Park, PA 16802, USA}
\author{S.~Campana}
\affiliation{INAF -- Osservatorio Astronomico di Brera, Via Bianchi 46, I-23807 Merate, Italy}
\author{G.~Cusumano}
\affiliation{INAF -- IASF Palermo, via Ugo La Malfa 153, I-90146, Palermo, Italy}
\author{A.~D'A\`\i}
\affiliation{INAF -- IASF Palermo, via Ugo La Malfa 153, I-90146, Palermo, Italy}
\author{P.~D'Avanzo}
\affiliation{INAF -- Osservatorio Astronomico di Brera, Via Bianchi 46, I-23807 Merate, Italy}
\author{V.~D'Elia}
\affiliation{INAF-Osservatorio Astronomico di Roma, via Frascati 33, I-00040 Monteporzio Catone, Italy}
\affiliation{Space Science Data Center (SSDC) - Agenzia Spaziale Italiana (ASI), I-00133 Roma, Italy }
\author{M.~de~Pasquale}
\affiliation{Department of Astronomy and Space Sciences, Istanbul University, Beyz{\i}t 34119, Istanbul, Turkey}
\author{S.~W.~K.~Emery}
\affiliation{University College London, Mullard Space Science Laboratory, Holmbury St. Mary, Dorking, RH5 6NT, U.K.}
\author{J.~Garcia}
\affiliation{Cahill Center for Astronomy and Astrophysics, California Institute of Technology, 1200 East California Boulevard, Pasadena, CA 91125, USA}
\author{P.~Giommi}
\affiliation{Space Science Data Center (SSDC) - Agenzia Spaziale Italiana (ASI), I-00133 Roma, Italy }
\author{C.~Gronwall}
\affiliation{Department of Astronomy and Astrophysics, The Pennsylvania State University, 525 Davey Laboratory, University Park, PA 16802, USA}
\affiliation{Institute for Gravitation and the Cosmos, The Pennsylvania State University, University Park, PA 16802}
\author{D.~H.~Hartmann}
\affiliation{Department of Physics and Astronomy, Clemson University, Kinard Lab of Physics, USA}
\author{H.~A.~Krimm}
\affiliation{National Science Foundation, Alexandria, VA 22314, USA}
\author{N.~P.~M.~Kuin}
\affiliation{University College London, Mullard Space Science Laboratory, Holmbury St. Mary, Dorking, RH5 6NT, U.K.}
\author{A.~Lien}
\affiliation{Center for Research and Exploration in Space Science and Technology (CRESST) and NASA Goddard Space Flight Center, Greenbelt MD, 20771 USA}
\affiliation{Department of Physics, University of Maryland, Baltimore County, 1000 Hilltop Circle, Baltimore, MD 21250, USA}
\author{D.~B.~Malesani}
\affiliation{Dark Cosmology Centre, Niels Bohr Institute, University of Copenhagen, Juliane Maries Vej 30, DK-2100 Copenhagen \O, Denmark}
\affiliation{DTU Space, National Space Institute, Technical University of Denmark, Elektrovej 327, 2800 Kongens Lyngby, Denmark}
\author{F.~E.~Marshall}
\affiliation{Astrophysics Science Division, NASA Goddard Space Flight Center, Greenbelt MD, 20771 USA}
\author{A.~Melandri}
\affiliation{INAF -- Osservatorio Astronomico di Brera, Via Bianchi 46, I-23807 Merate, Italy}
\author{J.~A.~Nousek}
\affiliation{Department of Astronomy and Astrophysics, The Pennsylvania State University, University Park, PA 16802, USA}
\author{S.~R.~Oates}
\affiliation{Department of Physics, University of Warwick, Coventry, CV4 7AL, UK}
\author{P.~T.~O'Brien}
\affiliation{Department of Physics and Astronomy, University of Leicester, University Road, Leicester, LE1 7RH, UK}
\author[0000-0002-1041-7542]{J.~P.~Osborne}
\affiliation{Department of Physics and Astronomy, University of Leicester, University Road, Leicester, LE1 7RH, UK}
\author[0000-0001-5624-2613]{K.~L.~Page}
\affiliation{Department of Physics and Astronomy, University of Leicester, University Road, Leicester, LE1 7RH, UK}
\author[0000-0001-7128-0802]{D.~M.~Palmer}
\affiliation{Los Alamos National Laboratory, B244, Los Alamos, NM, 87545, USA }
\author{M.~Perri}
\affiliation{Space Science Data Center (SSDC) - Agenzia Spaziale Italiana (ASI), I-00133 Roma, Italy }
\affiliation{INAF-Osservatorio Astronomico di Roma, via Frascati 33, I-00040 Monteporzio Catone, Italy}
\author{J.~L.~Racusin}
\affiliation{Astrophysics Science Division, NASA Goddard Space Flight Center, Greenbelt MD, 20771 USA}
\author{M.~H.~Siegel}
\affiliation{Department of Astronomy and Astrophysics, The Pennsylvania State University, University Park, PA 16802, USA}
\author{T.~Sakamoto}
\affiliation{Department of Physics and Mathematics, Aoyama Gakuin University, Sagamihara, Kanagawa, 252-5258, Japan}
\author{B.~Sbarufatti}
\affiliation{Department of Astronomy and Astrophysics, The Pennsylvania State University, University Park, PA 16802, USA}
\author{G.~Tagliaferri}
\affiliation{INAF -- Osservatorio Astronomico di Brera, Via Bianchi 46, I-23807 Merate, Italy}
\author{E.~Troja}
\affiliation{Astrophysics Science Division, NASA Goddard Space Flight Center, Greenbelt MD, 20771 USA}
\affiliation{Department of Physics and Astronomy, University of Maryland, College Park, MD 20742-4111, USA}

\begin{abstract}

The Neil Gehrels \swift\ Observatory carried out prompt searches for gravitational wave (GW) events detected by the LIGO/Virgo Collaboration (LVC) during the second observing run (``O2''). 
\swift\ performed extensive tiling of eight LVC triggers, two of which had very low false-alarm rates (GW 170814 and the epochal GW 170817), indicating a high confidence of being astrophysical in origin; the latter was the first GW event to have an electromagnetic counterpart detected.
In this paper we describe the follow-up performed during O2 and the results of our searches. 
No GW electromagnetic counterparts were detected; this result is expected, as GW 170817 remained the only astrophysical event containing at least one neutron star after LVC's later retraction of some events.
A number of X-ray sources were detected, with the majority of identified sources being active galactic nuclei.
We discuss the detection rate of transient X-ray sources and their implications in the O2 tiling searches. 
Finally, we describe the lessons learned during O2, and how these are being used to improve the \swift\ follow-up of GW events.
In particular, we simulate a population of GRB afterglows to evaluate our source ranking system's ability to differentiate them from unrelated and uncatalogued X-ray sources.
We find that $\approx60-70\%$ of afterglows whose jets are oriented towards Earth will be given high rank (i.e., ``interesting'' designation) by the completion of our second follow-up phase (assuming their location in the sky was observed), but that this fraction can be increased to nearly 100\% by performing a third follow-up observation of sources exhibiting fading behavior.

\end{abstract}


\section{Introduction} 
\label{sec:intro}

In 2017, the Advanced Laser Interferometer Gravitational-wave Observatory (aLIGO; \citealt{LIGO2015}) and the Advanced Virgo detector (the Virgo Scientific Collaboration; \citealt{VIRGO2015}) collectively carried out the second observing run (``O2'') in search of gravitational wave (GW) events from 30 November 2016 to 25 August 2017. 
The GW triggers detected by the LIGO-Virgo Collaboration (LVC) were assigned parameters including a false alarm rate (FAR; characterizing the frequency at which noise with the same strength as the signal is expected to arise), whether the detected signal arose from a compact binary coalescence (CBC) or an unmodeled burst\footnote{See https://www.ligo.org/science/GW-Burst.php}, and (for CBC triggers) the estimated distance of the merger and the masses of the initial compact objects. 
Triggers with a FAR of less than one per month were announced to electromagnetic (EM) follow-up partners who had signed a memorandum of understanding with the LVC.
O2 resulted in the detection of GW 170817, a binary neutron star (BNS) merger, which was the first GW event to have its electromagnetic counterpart (AT 2017gfo) detected \citep{Abbott2017}.
The results of both O1 and O2 are summarized in the Gravitational-Wave Transient Catalog of Compact Binary Mergers (GWTC-1; \citealt{LVC2018}).

In addition to GWs, electromagnetic radiation is expected to be produced in both binary neutron star (BNS) and neutron star black hole (NSBH) mergers, as was demonstrated in the case of GW 170817 (at least for the BNS case).
If the Earth lies close to the axis of rotation of the compact objects (i.e., ``on axis''), the prompt emission from the resulting relativistic jet is expected to be visible as a short gamma-ray burst (sGRB; see, e.g., \citealt{Berger2014}, \citealt{DAvanzo2015}, and \citealt{Beniamini2019}). 
On longer timescales, the radioactive decay of heavy r-process nuclei can produce broadband EM radiation visible as a kilonova, regardless of the viewing angle (\citealt{Eichler1989}, \citealt{Li1998}, \citealt{Metzger2010}).
Binary black hole (BBH) mergers are not typically expected to produce EM radiation \citep{Kamble2013}. 
However, it is theorized that under certain circumstances and with particular BH parameters (e.g., charged black holes, or if accreting or circumstellar material is present) BBH mergers may be able to produce EM radiation (see, e.g., \citealt{Loeb2016}, \citealt{Perna2016}, \citealt{Yamazaki2016}, \citealt{Zhang2016}, \citealt{Liu2016}), though this has yet to be observationally verified.
Thus, to further our understanding of the physics of compact binary mergers, it is necessary to search for and study the EM counterparts to merger events following the detection of their GWs.

\subsection{The Neil Gehrels \swift\ Observatory}
The Neil Gehrels \swift\ Observatory \citep{Gehrels2004} is a multiwavelength space-based NASA observatory whose primary mission is to detect and study GRBs and their afterglows in (soft) $\gamma$-rays, X-rays, ultraviolet, and optical wavelengths.
The Burst Alert Telescope (BAT; \citealt{Barthelmy2005}) is designed to detect GRBs in the 15--350 keV range using a coded aperture mask which covers a $\sim$2 sr field of view (FOV). 
The X-ray Telescope (XRT; \citealt{Burrows2005}) is an imaging instrument operating in the 0.3--10 keV range, with a circular 23.6$'$-diameter FOV.
The Ultraviolet/Optical Telescope (UVOT; \citealt{Roming2005}) covers the 1600--6240 \AA\ band with six filters, and the 1600--8000 \AA\ band with a white filter.
Its FOV is square, with sides of $\sim$17 arcminutes. 
Upon detecting a GRB, the BAT obtains its position (usually to within an accuracy of 1--4 arcminutes), and the spacecraft autonomously slews to the GRB's position within minutes (if there are no observing constraints).
The XRT and UVOT then observe the GRB afterglow and obtain arcsecond-scale localizations.

As mentioned by \citet{Evans2016c}, in an ideal scenario, the BAT would detect and localize the sGRB produced by a binary merger event independently of the detection of GW waves, promptly slew to the source, and detect the afterglow.
However, GRBs are only seen if the Earth lies within the jet's opening angle\footnote{The detection of prompt emission from sGRB 170817A by the \fermi-GBM and and {\sl INTEGRAL}-SPI, despite being off-axis by $\sim20^\circ$ \citep{Ghirlanda2019}, adds complications to this assumption by suggesting that at least some GRB jets may be structured and may release fainter X-ray emission over a wider angle. The prompt emission of sGRB 170817 was not seen by the BAT as occurred outside the BAT's field of view at the time.}$^,$\footnote{As well as GRB jet emission, there may also be more-isotropic X-ray emission (see, e.g., \citealt{Sun2017}).} 
The opening angles of GRB jets are not well measured.
Their model-predicted values vary, with estimates currently placing the jet opening angle $\theta_j \approx 3-10^\circ$, though some estimates are as high as $\theta_j\approx20^\circ$ (see, e.g., \citealt{Berger2014}, \citealt{Bloom2001}, \citealt{Frail2001}, and \citealt{Berger2003}). 
For a randomized distribution of sGRB jet axis orientations, the opening angles of $3^\circ$/$10^\circ$/$20^\circ$ correspond to 0.034\%/0.38\%/1.5\% of sGRBs occurring on-axis (i.e, that is the fraction of the sky that would be encircled by both jet and counter-jet for the above angles). 
Conversely, the gravitational waves emitted by merger events, though not isotropic, have a much weaker dependence on angle (the angular dependence is $\propto\cos(\theta/2)$). 
Therefore, only a small fraction of detected BNS/NSBH merger events will produce sGRBs that are visible to Earth.
Combined with the BAT's limited field of view (roughly 1/6 of the sky at any time), simultaneous LVC--BAT detections are expected to be uncommon.

In the event of a GW trigger, \swift\ can cover substantial portions of the GW error region with the XRT and UVOT in relatively short amounts of time (see \citealt{Evans2016c} for a discussion of the \swift\ follow-up to O1 triggers). 
Since the all-sky transient rate in X-rays (at \swift-XRT's sensitivity) is lower than that in the UV/optical bands (at UVOT's sensitivity; see, e.g., \citealt{Kanner2013,Evans2016a}), and with the XRT's FOV being larger than UVOT's, \swift's XRT also plays an important role in the search for and identification of EM counterparts to GW events. 
Since sGRB afterglows are not the only type of X-ray transients, as we search large areas of the sky it is important to consider the possibility of coincidental detections of unrelated X-ray transients and to quantify the rates of unrelated source detections in GW follow-up searches. 
The goal of this paper is to investigate and report on the rate of X-ray transients detected by \swift\ during O2; in particular, in the context of those detectable during the exposure times and timescales of the \swift\ GW follow-up procedures.

The paper is structured as follows.
In Section 2 we present an overview of the general observing strategy used, and information about the LVC GW triggers which we followed up.
In Section 3 we describe the data analysis techniques and source detection/flagging algorithms used.
In Section 4 we present the results of the searches and properties of the population of detected sources.
In Section 5 we discuss the implications of our results in the context of \swift\ follow-up of GW triggers in the Advanced LIGO/Virgo O3 run.

\section{Swift Response to GW events}

\subsection{Follow-up Observing Strategy}

The observing strategy employed in this campaign is the same as that which has been described in detail by \citet{Evans2016a,Evans2016c}, so only a brief summary is provided here. 

GW positional error regions can often encompass areas up to hundreds of deg$^2$.
Since the XRT field of view is only $23.6'$ in diameter, many pointings (tilings) are required to cover even a fraction of the higher-probability areas of the error region. 
In most cases covering the entire region within a reasonable time frame is not even feasible.
CBCs are believed to occur in or near galaxies, (see, e.g., \citealt{Fong2010}; \citealt{Tunnicliffe2014}).
Hence, a logical follow-up strategy is to convolve the LVC probability map and estimated distance of the triggers with the appropriate galaxy catalog (see Section 3.2 of \citealt{Evans2016c} and also \citealt{Evans2019} for details). 
This method reduces the area that needs to be observed for each trigger by focusing on fields containing known galaxies which are possible hosts to the merger event.
We used two catalogs: the 2MASS Photometric Redshift Catalog (2MPZ; \citealt{Bilicki2014}) and the Gravitational Wave Galaxy Catalogue (GWGC; \citealt{White2011}).
For GW events where the mean estimated distance was $\leq 80$ Mpc, we used the GWGC because it is more complete than 2MPZ in this regime\footnote{We also used this catalog for unmodeled burst events, which are only expected to be detectable within 100 Mpc \citep{LVC2019}.}.
For the more distant events, 2MPZ was used since is more complete.
The selected galaxies (i.e., located within the LVC region and consistent with the estimated distance to the GW event) were then prioritized based on distance and luminosity (which is used as a proxy for mass), as the latter is expected to be an indicator of sGRB rate (see \citealt{Fong2013}). 
When using the GWGC, galaxies were weighted by their $B$-band luminosity, and when using 2MPZ, the $K$-band was used (i.e., the native bands of the catalogs; the impact of this is investigated in Figure 7 of \citealt{Evans2016c}). 
The (in)completeness of the catalogs are included in the convolution procedure (c.f.\ equations 5, 6, and 10 in the above paper).
It is worth noting the caveat that there is a non-zero probability that the ``correct'' field will not be observed, due to the fact that we are not be able to cover anything near 100\% of the GW probability region even after galaxy convolution.
However, simulations by \citet{Evans2016a} show that the galaxy catalog convolution / targeted search method is more effective at detecting the EM counterpart before it fades than blindly searching the entire raw GW error region, which is far more time consuming.

The overall observing strategy was to carry out three phases of observations. 
First (phase 1), a series of short (60 s) exposures were taken covering as much of the galaxy-map-convolved GW error region as possible. 
This phase was designed with the intention of detecting an on-axis sGRB afterglow, if present, and continued for $\sim$2 days. 
Next (phase 2), 2--3 days after the trigger, the GW error region was re-observed for 500 s per tile. 
These observations, which continued for up to 4 days, were optimized to search for the rising X-ray afterglow from an sGRB observed off-axis. 
Collectively, phases 1 and 2 are referred to as the ``wide-area search'' phase.
Finally (phase 3), if no confirmed counterpart was found, any potentially interesting (i.e., unidentified) X-ray source was re-observed with exposures $>$1 ks. 
Any such source found to be fading was re-observed repeatedly over the following days until it could be confirmed to be the counterpart, or ruled out as such.

In reality, this program was not always followed completely. 
The prolonged period of many short-exposure observations was unlike any previous use of \swift, and so the number of fields observed for the first few triggers were reduced to allow us to verify that this observing mode did not pose a risk to the spacecraft, as the large number of slews carried out in such a short time period was unprecedented. 
Even once this had been confirmed, we only carried out all three phases fully in one case, that of G275697 (see Table \ref{tab:obs_overview}). 
For the BBH triggers, where an EM counterpart is not expected, we only carried out phase 1. 
Phase 2 was carried out for all BNS triggers except G275404 (because trigger G275697 occured when phase 2 was due to start and we decided to prioritize the more-nearby event) and G298048 (GW 170817, for which the real counterpart was found and therefore phase 2 was unnecessary). 
For trigger G299232, only phases 1 and 2 we carried out. 
Additionally, we occasionally carried out targeted observations of potential counterparts reported in the GCN Circulars by other facilities.

\subsection{Follow-up Criteria}
GW trigger notices were issued by the LVC for any event with a FAR of $<1$ per month.
These trigger notices also included an estimated $P_{\rm NS}$, the probability that the event involved at least one neutron star.
We evaluated each trigger (convolved the sky map, estimated distance, and appropriate galaxy catalogs) and assigned each a ``P400'' value, which is the fraction of the probability region contained within the 400 most probable XRT fields taken from the galaxy-convolved skymap that were not Sun- or Moon-constrained by \swift\footnote{A target is Sun/Moon-constrained if it lies within 47$^\circ$/23$^\circ$ of the Sun/Moon, respectively.}.
This essentially quantifies how much of the LVC probability region \swift-XRT can cover within one day.

The decision tree was set the following way: \swift\ would follow-up an event only in the following cases:
\begin{itemize}
\item For burst (unmodeled) triggers:
{\indentitem \item if FAR $<1/6$ month$^{-1}$ and P400 $>0.2$}
\item For CBC triggers:
{\indentitem \item if $P_{\rm NS} < 0.25$ and P400 $>0.5$}
{\indentitem \item if $P_{\rm NS} > 0.25$ (regardless of P400)}
\end{itemize}

To test our follow-up response protocol, it was predetermined that the first two events in O2 would be followed up regardless of their qualifying criteria.

\section{Data Analysis}
The XRT data were automatically processed at the United Kingdom \swift\ Data Science Centre (UKSSDC) at the University of Leicester, using {\tt HEASOFT} v6.22 and the latest {\tt CALDB} available at the time of processing.
Observation data were initially reprocessed using the {\tt XRTPIPELINE} tool, which applied all necessary calibrations, filtering, and corrections\footnote{For more details, see http://www.swift.ac.uk/analysis/ and https://heasarc.gsfc.nasa.gov/ftools/caldb/help/xrtpipeline.html}.
Images and exposure maps of each observation were also created.

The basic steps of the XRT analysis of GW follow-up are as follows: (1) search for sources, (2) characterize sources, and (3) identify any potential counterparts to the GW trigger.
Initial source detection was carried out during the wide-area search phase of the follow-up with the goal of finding sources of interest.
The wide-area search phase includes the initial 60 s and 500 s exposures of each field.  
The source detection procedure is an iterative process which involves sliding-cell source detection, background modeling, PSF-fitting, and a likelihood test to detect and localize the sources.
This method was the same as that which has been used to produce the \swift\ X-Ray Point Source Catalog (1SXPS); it was described in detail by \citet{Evans2014}. 
The pipeline assigned each detected source a quality flag, which characterizes the probability of the source being a spurious detection.
Sources flagged as ``good'' have a 0.3\% or less chance of being spurious (or false positive; FP), ``reasonable'' sources have a 7\% FP rate, and ``poor'' sources have up a 35\% FP rate.
Considering both ``good'' and ``reasonable'' sources together yields a 1\% FP rate (as ``good'' sources are the most numerous), and all ``good'', ``reasonable'', and ``poor'' taken together result in an overall FP rate of roughly 10\%. 
The detected sources are manually verified for spurious detections that can arise from optical loading, stray light, extended emission, and/or thermal noise (which can result from the XRT detector getting too hot).

As previously noted by \citet{Metzger2012}, when following up on LVC triggers, the major challenge is not only detecting an EM candidate counterpart but also discerning which, among the many sources detected, if any, is the actual EM counterpart to the GW event.
\citet{Evans2015} discussed two methods of discerning an X-ray GRB afterglow from unrelated sources on the basis of either brightness and/or fading behavior.
A source can be a potential afterglow if it is bright enough that it should have been previously catalogued but has not been, in which case it has exhibited transient behavior.
Additionally, sGRB X-ray afterglows fade on relatively short timescales, so a source may be a counterpart candidate even if it is below a catalog limit if it is fading rapidly.
Therefore, the source characterization procedure is based off these considerations.
The process accounted for the source brightness (in comparison to historical detections and flux limits), light curve behavior, and whether the source lies within 200 kpc (in projection) of a known galaxy (see \citealt{Bulik1999}) with distance consistent (at the 3-$\sigma$ level) with the distance estimate along that line of sight from the GW data.
The flux limits were compared with those from the ROSAT All Sky Survey (RASS; \citealt{Voges1999}) and, where observations existed, \xmm\ observations (both pointed and slew surveys; \citealt{Saxton2008}) and the 1SXPS catalog \citep{Evans2014}.

Each source was placed into one of four rankings (as defined by \citealt{Evans2016b}), described below in decreasing order of importance.
 
Of highest priority (``rank 1'') are afterglow candidates.
Sources given this designation were either: (1) uncatalogued and at least $5\sigma$ above the $3\sigma$ upper limit from the RASS or 1SXPS, or (2) a known X-ray source which is $5\sigma$ above its catalogued flux\footnote{The historical count-rate/upper limits for both criteria were not derived from XRT data; they have been converted to equivalent XRT (PC mode) 0.3--10 keV count rates using {\tt PIMMS} (Portable Interactive Multi-Mission Simulator), assuming a typical AGN spectrum (absorbing Hydrogen column density $N_{\rm H}=3\times10^{20}$ cm$^{-2}$, and photon index $\Gamma=1.7$).  The peak source fluxes were also obtained by converting from the peak count rates when assuming a typical AGN spectrum with the above-mentioned parameters.}.
Afterglow candidates must also lie near (within 200 kpc in projection of) a known galaxy (assuming the source is at the distance of that galaxy).

Of subsequent importance are ``interesting'' sources (``rank 2'').
These are either: (1) uncatalogued and: at least $3\sigma$ above the $3\sigma$ upper limit from the RASS/1SXPS or fading, or (2) known X-ray sources at least 3$\sigma$ above their catalogued flux.
Unlike afterglow candidates, an interesting source need not be near a known galaxy.

Next are uncatalogued X-ray sources (``rank 3'').
These were objects which were not previously catalogued in X-rays, but also meet none of the above criteria to differentiate them from field sources unrelated to a GW trigger. 

Of least interest are known X-ray sources (``rank 4'').
This category includes objects which have been detected in X-rays before, and have a flux consistent with or below that from the previous observations. 
No further follow-up action is taken for these sources.

If a rank 1 source were detected at any point, the seach would have been interrupted so the rank 1 source could be re-observed immediately.
If the rank 1 source were determined not to be the afterglow, the search would be resumed. 
After the initial 60 and 500 s observations were completed, deeper follow-up observations were carried out (``phase 3'').
Rank 2 sources were re-observed with deeper exposures $\sim5-6$ ks. 
If no afterglow has been found at this point, rank 3 sources are then re-observed with 1 ks observations. 
Some selected fields and/or sources were followed-up for longer periods ($>10$ ks) due to target of opportunity (ToO) requests submitted by members of the astrophysical community. 
Phases 1-3 were only carried out in full for one trigger (G275697; see below).

\begin{table*}
\begin{center}
\caption{Overview of GW Triggers and \swift\ Follow-up in LVC O2}
\label{tab:obs_overview}
\begin{tabular}{ccccccccccccc}
\hline
Trigger ID &  Trigger date & $d_{\rm est}$ & $P_{\rm NS}$ & Delay & Duration  &  Exposure & Num.\ & Area &  $P_{\rm raw}$  & $P_{\rm conv}$  & $N_{\rm XRT}$  & $N_{\rm uncat}$ \\  
 & (2017; UT) & (Mpc) &  & (ks) & (ks) & (ks) & fields & (deg$^2$)  \\
\hline
G268556 & 01-04, 10:11:59 & $737\pm201$ & 0\% & 50.3 & 1174 & 37 & 293 & 31.7 & 4.7\% & 4.6\% & 3 & 0 \\ 
G270580 & 01-20, 12:31:00 & $\lesssim100$ & (b) & 19.9 & 351 & 10 & 136 & 14.5 & 1.1\% & 14\% & 2 & 0 \\ 
G274296 & 02-17, 06:05:53 & $\lesssim100$ & (b) & & & &  &  &  &  &  &  \\ 
G275404 & 02-25, 18:30:21 & $412\pm169$ & 100\% & 17.3 & 1121 & 12 & 117 & 2.7 & 1.8\% & 4.0\% & 1 & 0 \\ 
G275697$^\dagger$ & 02-27, 18:57:31 & $193\pm61$ &  100\% & 15.9 & 519 & 257 & 1408 & 171 & 16\% & 31\% & 58 & 16 \\ 
G277583 & 03-13, 22:40:09 & $\lesssim100$ & (b) & & & &  &  &  &  &  &  \\ 
G284239 & 05-02, 22:26:07 & $\lesssim100$ & (b) & & & &  &  &  &  &  &  \\ 
G288732 & 06-08, 02:01:16 & $320\pm98$ & 0\% & 62.8 & 29 & 9 & 4 & 0.5 & 0\% & 0\% & 0 & 0  \\ 
G296853 & 08-09, 08:28:21 & $1086\pm302$ & 0\% & & & &  &   &  &  &  &  \\ 
G297595 & 08-14, 10:30:43 & $534\pm131$ & 0\% & 31.2 & 517 & 113 & 643 & 68.0 & 24\% & 36\% & 41 & 15 \\ 
G298048 & 08-17, 12:41:04 & $39\pm7$ & 100\% & 3.3 & 9206 & 269 & 744 & 85.0 & 2.5\% & 94\% & 12 & 1  \\ 
G298389 & 08-19, 15:50:46 & $\lesssim100$ & (b) & & & &  &  &  &  &  &   \\ 
G298936 & 08-23, 13:13:58 & $1738\pm477$ & 0\% & & & &  &  &  &  &  &   \\ 
G299232 & 08-25, 13:13:37 & $339\pm109$ & 100\% & 11.1 & 748 & 156 & 653 & 75.9 & 8.3\% & 16\% & 40 & 19 \\ 
\hline
\end{tabular}
\end{center}
\begin{flushleft}
$d_{\rm est}$ is the estimated distance to the GW event in Mpc. 
$P_{\rm NS}$ is the probability that the event included at least one neutron star, for CBC triggers; ``(b)'' denotes that an event was a burst GW event, in which case (since little is known about their origin) the probability they involve a neutron star can not be determined. 
The delay is the interval between the GW trigger time and the time at which the first follow-up observation began. 
The duration is the time from the start of the first observation with \swift-XRT to the end of the last one. 
\swift-XRT was not observing the GW region for the entirety of this time, so the total exposure is given in the subsequent column.
The area listed is corrected for the overlaps between adjacent tiles. 
$P_{\rm raw}$ is the fraction of the LVC skymap which was enclosed by XRT observations, and $P_{\rm conv}$ is the fraction of the galaxy-convolved skymap which was covered by XRT observations.  
$N_{\rm XRT}$ is the number of sources detected by the XRT in each follow-up search (only phases 1 and 2), and $N_{\rm uncat}$ is the number (of $N_{\rm XRT}$) which are uncatalogued.
$^\dagger$ Trigger G275697 was retracted after the follow-up search was performed.
Triggers G275404 and G299232 (although originally marked with $P_{\rm NS}=100\%$) were later determined not to be real astronomical events (see \citealt{Abbott2019}).
For completeness, it is also worth noting that there were two additional GW triggers which were only detected in post-O2 analyses, and consequently could not have been followed-up within a reasonable amount of time: GWs 170729 and 170818 (see \citealt{LVC2018}).
\end{flushleft}
\end{table*}

\section{Results}

\subsection{Follow-up Summary}

During O2, \swift\ carried out a follow-up search of 7 CBC triggers (all of which had estimated distances $d_{\rm est} < 1$ Gpc), and 1 unmodeled burst.
In total, 3998 XRT fields were observed, covering an area of 449.3 deg$^2$ (accounting for overlapping regions in XRT tilings; see, e.g., Figure 1 of \citealt{Evans2015}) in 863 ks of observation time.
A brief summary of the LVC O2 triggers and \swift\ follow-up searches is presented in Table \ref{tab:obs_overview}. 

It is worth noting that LVC trigger G275697 was subsequently retracted. 
Thus, the follow-up searches for this trigger are unique in that it is the only trigger for which we can be certain that we did not detect any counterpart (or that there was no counterpart to detect), since no actual astrophysical event took place.

LVC trigger G288732 did not meet our trigger criteria, however, a Target of Opportunity (ToO) request to observe a transient {\sl Fermi}-LAT source spatially consistent with the GW error region was submitted, so \emph{Swift}-XRT and UVOT began observing at 2017 June 08 at 19:27:20 (17.4 hr after the GW trigger).
A four-point tiling was selected to cover the {\sl Fermi}-LAT error region. 
The observations continued for 29 ks, until 2017 June 09 at 03:34:21, and gathered 9 ks of observation data. 

\swift\ also followed up LVC trigger G298048. 
This trigger was a binary neutron star trigger, the famous GW 170817.
No early X-ray emission was detected, however, due to the detection of the optical transient AT2017gfo (the first detection of an electromagnetic counterpart to a GW event), it was subsequently found that this GRB was off-axis and the afterglow rose later (see, e.g., \citealt{Haggard2017}, \citealt{Margutti2017}, \citealt{Margutti2018}, \citealt{Troja2017}, \citealt{Troja2018}).
This late-rising X-ray emission was not detected by the XRT, despite more than 50 ks of exposure time; contamination from the X-ray emitting host galaxy, combined with the relatively large PSF of XRT (9$''$ half-energy width) made it impossible to make a solid detection of AT2017gfo (i.e., distinguish the source's emission from that of its host galaxy).

More detailed descriptions of each trigger and follow-up searches are given in Appendix A.

\subsection{Detected Sources}
157 sources were detected in the wide-area search phase. 
Details for each source are listed in Table \ref{tab:cattable} (which can be found after the References section).

4 were flagged as ``interesting'' (rank 2).
Among those, 3 were uncatalogued X-ray sources: all of which exhibited significant fading ($4.1<\sigma<4.3$), and only one of which was near\footnote{The galaxies are checked using the GWGC and 2MPZ catalog.  To be considered ``nearby'' a galaxy, a source must be within 200 kpc  of a galaxy (in projection, at that distance), assuming the source is as the distance of the galaxy.  Thus, a source can be marked as ``having 0 nearby galaxies'', but be coincident with a galaxy that is outside the range of estimated distances of the GW trigger, in which case it would be unrelated to the trigger.} at least one galaxy (in this case, an AGN).
Though the catalogued source did not exhibit any signs of fading, its peak flux had increased over its catalogued limit with a significance $>3\sigma$.

51 sources were uncatalogued (rank 3; i.e., they were not previously detected in X-rays).
35 of these lacked any nearby known galaxies within the range of distances compatible with their related GW triggers (though this designation does not mean a source can not be associated with a galaxy at a distance less than or greater than its respective GW trigger).
10 rank 3 sources were (or are positionally coincident with known) galaxies (or AGN candidates).
8 rank 3 sources exhibited evidence of fading: 7 of these were of low significance ($1.1<\sigma<1.7$), and only 1 was of high significance (5.2$\sigma$).

102 sources detected were previously-catalogued X-ray (rank 4) sources. 
Of those, 41 were not located near any known galaxy that was consistent with the estimated distance of the GW trigger.
40 of the rank 4 sources were identified as (or are positionally coincident with known) galaxies (or AGN candidates).
8 sources exhibited slight evidence of fading ($1.1<\sigma<1.4$).
Of the fading sources, 4 are AGNs (or AGN candidates), 1 is a star, 1 is an eclipsing binary, and the remaining 2 are unknown.

153 of the sources were of the ``good'' detection quality flag, and 4 were of the ``reasonable'' quality flag.
Of the latter, 3 were rank 3 (and unidentified) sources, and 1 was a rank 4 (known X-ray) source.

\section{Discussion}

\subsection{What We Found} 

\begin{figure*}
\epsscale{1.1}
\plotone{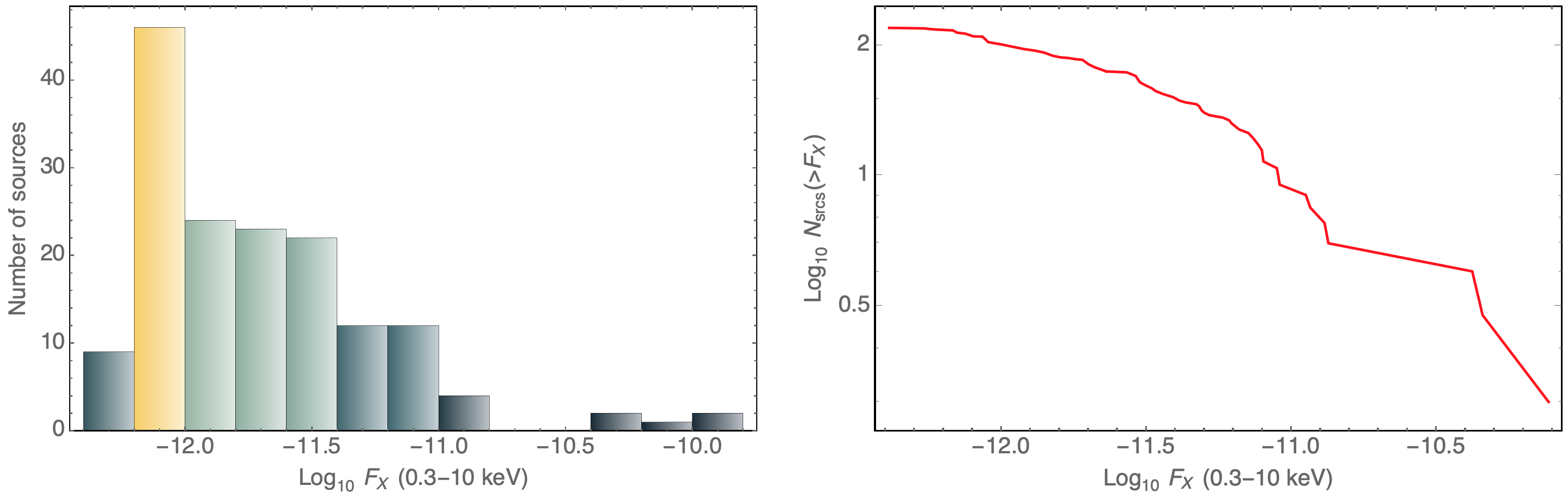}
\caption{{\sl Left:} Histogram of the X-ray fluxes $F_X$ of sources detected in the wide-area search phase (phases 1 and 2).
{\sl Right:} $\log N$ -- $\log S$ plot of the population of detected sources in the wide-area search phase. (Here we represent X-ray flux with $F_X$ instead of the traditional $S$).  Fluxes are in units of erg cm$^{-2}$ s$^{-1}$}.
\label{fig-histograms}
\end{figure*}

\begin{figure*}
\epsscale{1.1}
\plotone{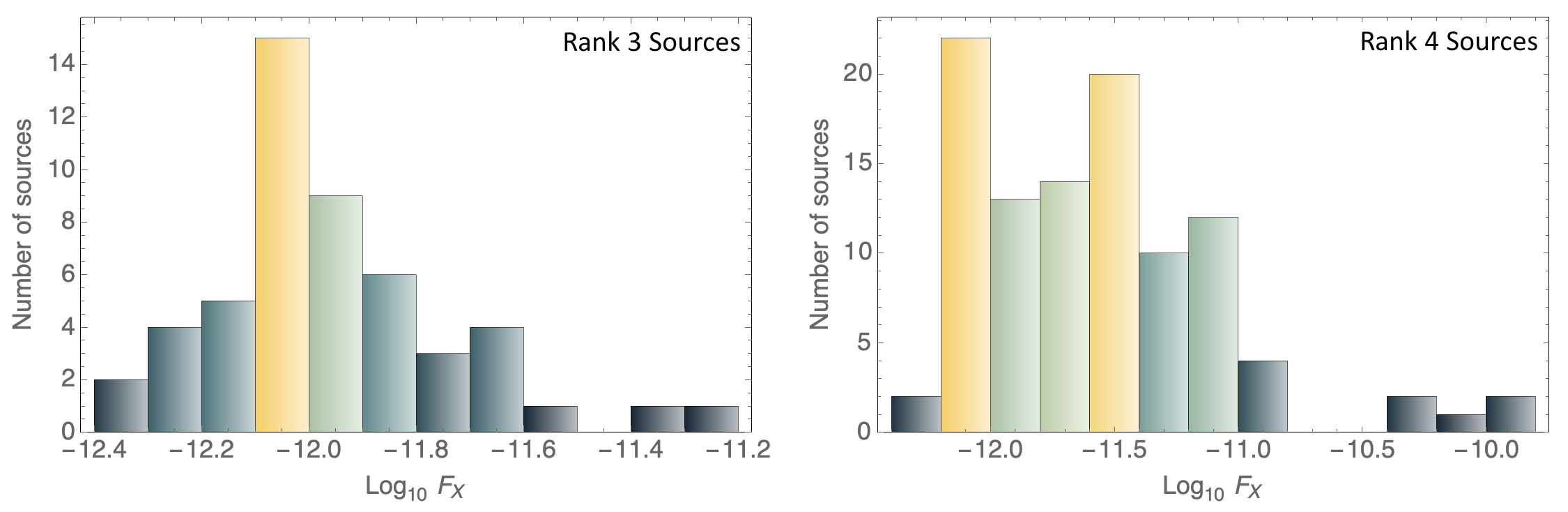}
\caption{Histograms of the X-ray (0.3--10 keV) fluxes $F_X$ of sources detected in the wide-search phase (phases 1 and 2) separated by source rank; left: rank 3, right: rank 4.  Fluxes are in units of erg cm$^{-2}$ s$^{-1}$.}
\label{fig-histograms-rank}
\end{figure*}

\begin{figure*}
\epsscale{1.0}
\plotone{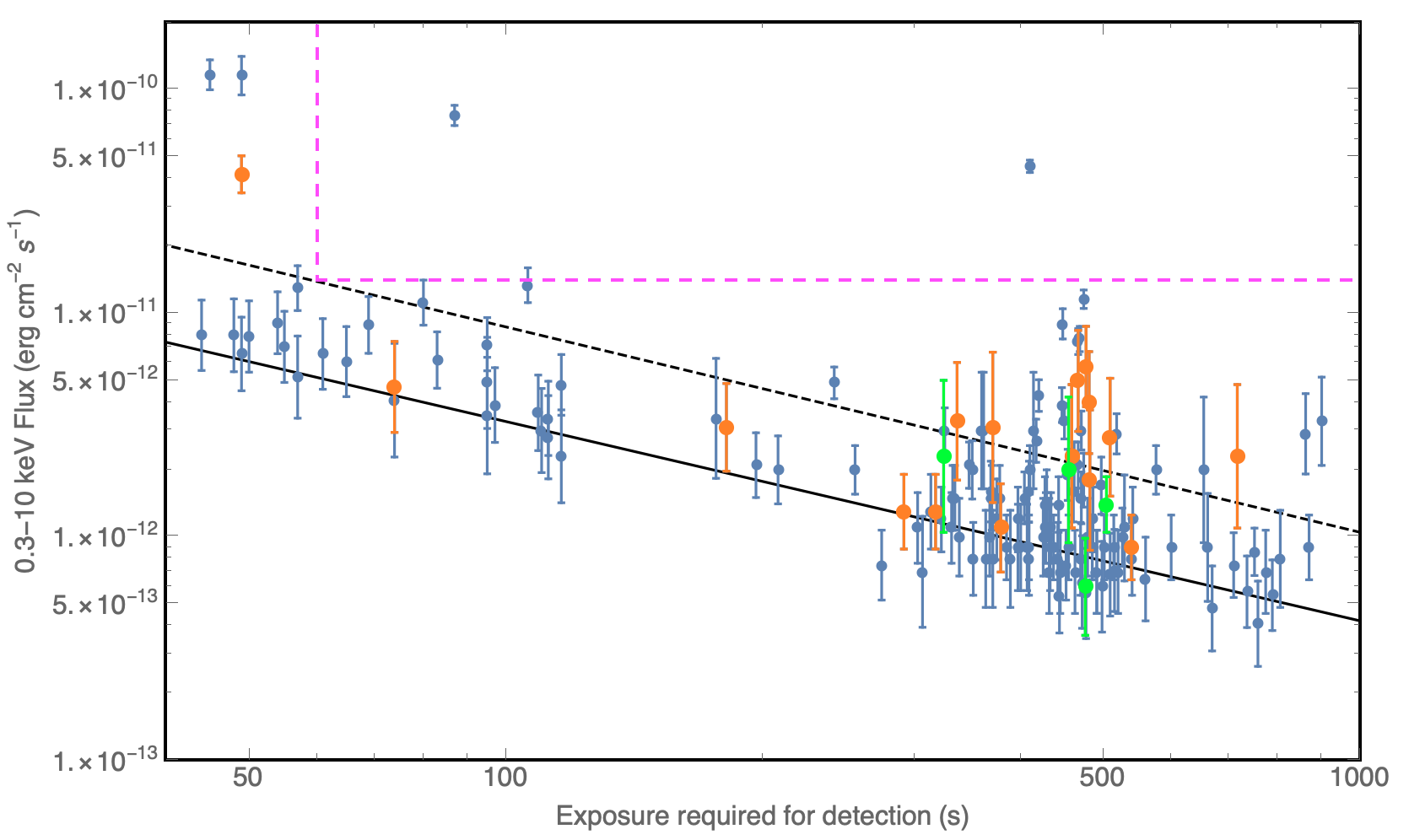}
\caption{Plot of the peak fluxes of all sources versus the theoretical exposure time required to detect each source. The black solid/dashed lines represent the XRT sensitivity as a function of exposure time, at the 50\%/90\% confidence levels (from 1SXPS), respectively (note that the sensitivity ``lines'' are rather curves, since they are shown in log space).  
The plotted points represent the peak fluxes of each source (with their corresponding 1$\sigma$ uncertainties; y-axis) versus the effective exposure time needed to detect a source at its flux.  
Sources which exhibited slight fading ($1<\sigma\leq2$) are highlighted in orange, and sources which exhibited more significant fading (at least $>2\sigma$) are highlighted in green (though these sources all have $>4\sigma$ significance).  
Some sources have exposures for detection $>560$ s, which were found in the overlapping regions of adjacent tilings.
The two sources fully enclosed within the dashed pink dashed lines are sources {\sl not} detected in the initial 60 s exposures, but should have been had they been this bright (i.e., at their peak flux) when those observations were carried out.
\vspace{0.2cm}}
\label{fig-sensitivity}
\end{figure*}

In Figure \ref{fig-histograms} we present a histogram of the peak fluxes of the detected sources (left panel) and a $\log N - \log S$ plot (right panel).
In Figure \ref{fig-histograms-rank} we present a histogram of the peak fluxes, but separated by source rank (ranks 3 and 4 are shown; no rank 1 sources and only four rank 2 sources were detected).
In Figure \ref{fig-sensitivity} we plot the peak fluxes of all sources versus the theoretical exposure time required to detect each source.
From these, we can see that few sources with peak fluxes $\lesssim6\times10^{-13}$ erg cm$^{-2}$ s$^{-1}$ were detected.

There are two sources which should have been detected in the initial 60 s exposures with $\approx$100\% confidence, but weren't (those that are enclosed within the dashed pink region).
If 90\% confidence is considered (i.e., those slightly below the horizontal pink line), then there are a couple more sources of this type.
This indicates that these sources have risen in brightness considerably between the initial exposure (phase 1; 60 s) and second deeper exposure (phase 2; 500 s) in timescales of $<5$ days.
These two prominent sources (i.e., within the pink region), 1SXPS J133553.6--341744 (trigger G298048, rank 4) and 3XMM J023819.7--521132 (trigger G297595, rank 4), are close to ($<5''$ from) ESO 383--35 and ESO 198--24, respectively, which are both Type 1 Seyfert galaxies (active galactic nuclei; AGN).

This has two implications.
The first is that a ``late'' detection of a bright source can be an indicator of variability, and could be used to identify potential GW counterparts (i.e., sGRBs whose jets were initially not on-axis but which later widened).
The second is that, the spacing of our 60 s and 500 s follow-up observations is consistent with the timescales of AGN variability, and that we are able to detect such behavior. 
Indeed, in X-rays, AGN are known to be able to vary down to timescales on the order of an hour (see, e.g., \citealt{Middei2017}). 
Since typically the area searched lies outside the Galactic plane (as sGRBs/GW events have an isotropic distribution in the sky), it is reasonable to assume that the majority of X-ray sources seen will be AGNs.
More than half (7/10) of the catalogued X-ray sources detected in our follow-up exhibiting variability are identified as AGNs.
The other three of the remaining catalogued and identified sources are stars (one of which is XMMSL1 J114247.5--354904, the counterpart to V* V752 Cen / HD 101799: an eclipsing binary with an 8-hour period; \citealt{Sistero1974}). 
Thus, a point of consideration in future GW follow-up campaigns is that if a GW counterpart occurred in an X-ray-active galaxy, the automated source ranking system would classify it as a known source getting brighter.
Since AGN commonly vary on timescales comparable to our follow-ups, differentiating between an AGN ``hiccup'' and a transient occurring in a galaxy with an active nucleus will be challenging.

About $13\% \pm3\%$ of the sources detected in the wide-area search phase (21/157) exhibited fading behavior with a significance $\geq1\sigma$.
It is worth noting that, assuming Gaussian errors, we would expect 16\% of constant sources to be identified as fading at a level $\geq1\sigma$; therefore, the observed fraction of fading sources is consistent with what is expected from a population of constant sources.
Two uncatalogued sources with fluxes slightly above the RASS limit were detected.
The flux and uncertainty of the first source (trigger G275697, rank 3), \swift\ J213954.9+444551.1, places it at $0.9\sigma$ above the RASS limit, however it exhibited fading behavior with $1.7\sigma$.
The second source (trigger G298048, rank 3), \swift\ J132507.3--323814.4, at $1.9\sigma$ above the RASS limit, is coincident with 2MASX J13250705--3238129 and the radio source VBM97 J1325--3238, which are coincident with the galaxy cluster Abell A3556, and thus this source is likely an AGN.
Considering the 2 most significant rising sources (those above the pink line in Figure \ref{fig-sensitivity}), the two above the RASS limit, and the catalogued interesting source, the fraction of sources exhibiting transient behavior in the wide-area search phase is $17\% \pm 3\%$ (26/157).

Although \swift\ did not find any counterparts, except for trigger G298048 (GW 170817), \swift\ was not expected to find any, as none of the other triggers in O2 contained a BNS merger and as triggers G275404 and G299232 were subsequently determined not to be real \citep{Abbott2019}, and as (for most triggers), only a small fraction of the probability regions were covered.

\begin{figure}
\epsscale{1.2}
\plotone{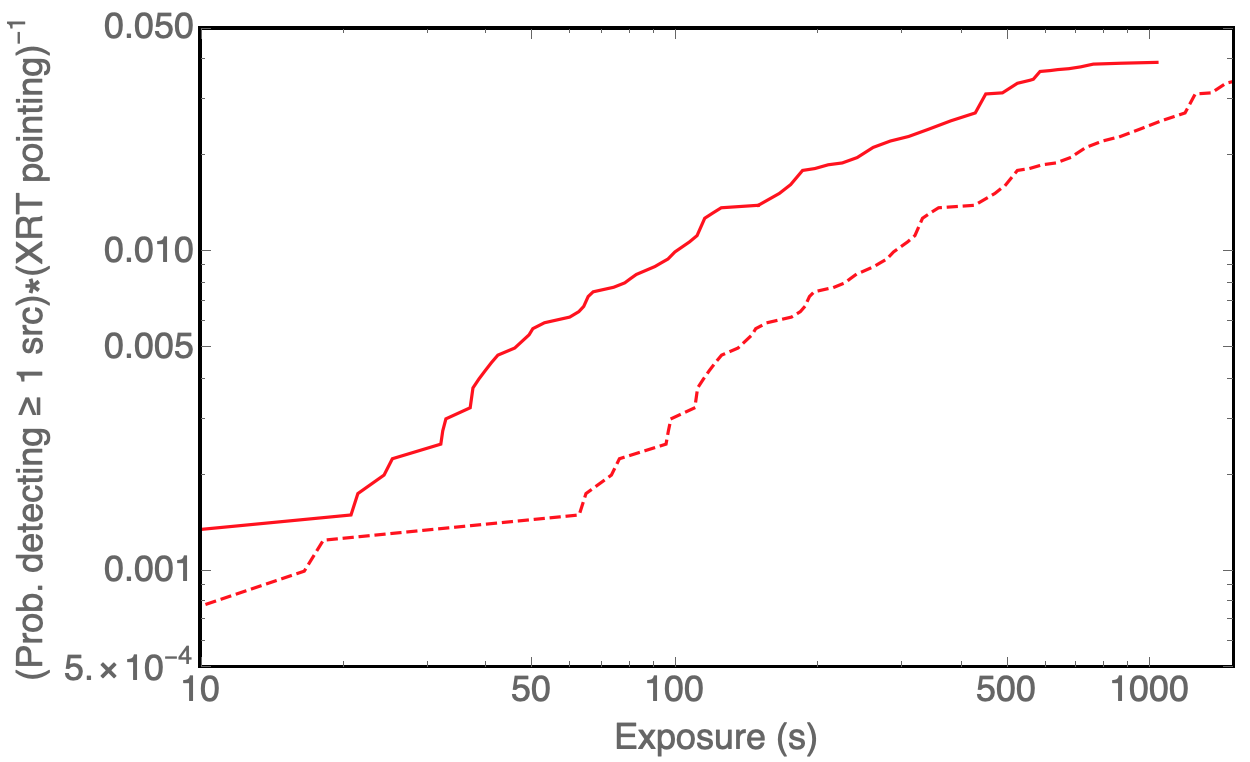}
\caption{Plot of the probability of detecting at least one source per XRT field of view as a function of exposure time.  The solid curve represents detections at 50\% confidence, and the dashed curve represents detections at 90\% confidence.}
\label{fig-detection-prob}
\end{figure}

\subsection{Looking Forward: What We Can Expect}
Using the peak fluxes of all 157 sources detected and the XRT sensitivity curves (Figure \ref{fig-sensitivity}) we can calculate the theoretical exposure times needed to detect each source at 50\% and 90\% confidence\footnote{In other words, the exposure times corresponding to a 50\% and 90\% chance of detecting a source (assuming the source's count rate can be described by a Poissonian process).}, and from that, construct a cumulative distribution function (CDF) for the probability of detecting a source as a function of exposure\footnote{It is worth noting that the sensitiviy at short exposures may not extrapolate perfectly.}. 
We divide the CDF by the total area covered in the GW searches (449.3 deg$^2$), and scale it to the area covered by the XRT field of view (FOV; 0.1215 deg$^2$). 
In Figure \ref{fig-detection-prob} we plot this CDF, which shows the probability of detecting at least one source (per unit area corresponding to the XRT FOV) as a function of exposure time.
After a 560 s exposure, the probabilities of detecting at least one source at 50\% and 90\% confidences in an XRT FOV are approximately 0.035 and 0.018, respectively.
Therefore, given the low rate of serendipitous X-ray sources expected, the probability of the GW counterpart being blended with an unrelated source is low (assuming that the counterpart is not in an X-ray-active host).

\subsubsection{Source Detection Rates}
Trigger G275697 was notable in that it was the first trigger for which phase 1 was carried out in full (after which, it is worth noting, no negative effects on the spacecraft from the large number of slews were observed).
It was also the only trigger for which phases 1, 2, and 3 were carried out in full.
Lastly and of most importance, as this trigger was retracted, it is the only follow-up search for which we can be certain that there was no actual afterglow.
Using this population of definitively-unrelated sources, we can place (crude) limits on the detection rates of serendipitous sources for each rank.
We can expect rank 1 and 2 sources to be detected at a rate of $<0.0175 \pm 0.0101$ per deg$^2$, rank 3 sources at a rate of $0.0936 \pm 0.0234$ per deg$^2$, and rank 4 sources at a rate of $0.2280 \pm 0.0365$ per deg$^2$.
If we also consider triggers G297595 (LVC 170814) and G299232 (LVC 170825), the other two triggers for which phase 2 observations were carried out, we can refine out estimates as follows\footnote{We do not consider triggers for which only phase 1 (60 s) observations were carried out, as \swift-XRT can not reach fluxes below than the RASS limit in such short exposures, and therefore we do not expect to detect uncatalogued sources.}:  we can expect to detect rank 1 and 2 sources at a rate of $<0.0127 \pm 0.0064$ per deg$^2$, rank 3 sources at a rate of $0.1588 \pm 0.0225$ per deg$^2$, and rank 4 sources at a rate of $0.2699 \pm 0.0293$ per deg$^2$ (assuming that none of the sources from triggers G297595 or G299232 were an actual counterpart).

\subsection{Evaluation of Source Ranking Criteria}
To investigate our efficacy in distinguishing an afterglow from unrelated X-ray sources (i.e., our ability to assign afterglows rank 1 or 2), we performed a simulation of afterglows and ran those with fluxes above our detection limits through our source ranking algorithm.
We used {\tt afterglowpy}\footnote{https://github.com/geoffryan/afterglowpy} \citep{Ryan2019}, a Python module that produces GRB light curves using the models of \citet{vanEerten2010} and \citet{vanEerten2018}, to simulate 10,000 sGRB events.

For all simulations, we used the typically-assumed electron thermal energy fraction $\epsilon_e=0.1$ (which is consistent with recent observational results; see, e.g, \citealt{Beniamini2017}) and $d=120$ Mpc (the expected average distance of BNS mergers in O3).
For the other parameters, we used the values obtained from X-ray observations of sGRBs (and those derived from observationally-determined parameters) from \citet{Fong2013}.
We randomly selected electron SED slope $p$ and jet energy $E_{\rm jet}$ from the list of observed values (Table 3 of \citealt{Fong2015}).
For the circum-burst number density $n$ we randomly selected a value between $(0.6,15)\times10^{-3}$ cm$^{-3}$, which corresponds to the best-fit value for GW/GRB 170817 (a rather low value) and the upper range of values considered typical, and for $\epsilon_B$ we selected a random value between 0.01 and 0.1 (thus sampling the range of typical values).
For the off-axis (viewing) angle $\theta_{\rm obs}$, we sampled from the expected distribution of off-angles detectable by LVC (see Equation 28 of \citealt{Schutz2011}).
For each simulated event, we calculated the 0.3-10 keV flux at two times corresponding to the typical elapsed times between the trigger, $t_0$, and our phase 1 and phase 2 observations, $t_1$, $t_2$.
For $t_1$ we picked a random time between $t_0 + 1.5$ hr (1.1 hr for the average time between ground station passes during which we can upload tiling plans, and 0.4 hr for roughly half of a \swift\ orbit) and $t_0 + 1.5$ d $+ 1.5$ hr (i.e., within the first half of phase 1; we assume the afterglow will lie in the higher-probability areas which we observe first).
For $t_2$ we picked a random time between $t_0 + 3$ d $+ 1.5$ hr (the start of phase 2) and $t_0 + 4$ d $+ 1.5$ hr (the second day of phase 2).
If the afterglow's flux at $t_1$ and/or $t_2$ reaches our detection thresholds for 80 s and/or 500 s tilings (the exposure times we will use in O3 follow-up; thresholds are $4.5\times10^{-12}$ and $8\times10^{-13}$ erg s$^{-1}$ cm$^{-2}$, respectively), we simulated the observed XRT count rate in {\tt PIMMS} using that sGRB's flux, power-law slope $p$, and an $N_{\rm H}$ randomly-selected between $10^{21}$ and $10^{22}$ cm$^{-2}$ (the approximate range of typical values; see Figure 7a of \citealt{Evans2009}\footnote{see https://www.swift.ac.uk/xrt\_live\_cat/\#figureDiv}).
Our source ranking procedure (described above) takes into account the RASS upper limits, which can vary by RA/Dec, so we obtained a ``typical'' RASS upper limit by taking the average of limits at 3 high-probability regions of each LVC trigger we performed tilings for.
We gathered these limits from the ESA Upper Limits Server\footnote{http://xmmuls.esac.esa.int/upperlimitserver/}, and converted them into an 0.3--10 keV XRT count rate (we found the average value to be 0.023 ct s$^{-1}$).
We tested the above simulations with the option for jet spreading turned off (with initial Lorentz factor $\gamma_0=1000$, the recommended value) and turned on (with initial Lorentz factor $\gamma_0$ set to infinity, the default value), and our results were the same (suggesting that  jet spreading does not have a noticeable effect on the observed fluxes over the elapsed times we used for $t_1$ and $t_2$).
For the jet, we assumed a Gaussian structure with a wing truncation angle of $20.1^\circ$ (the default value), and re-performed the entire simulation using each of the following opening angles $\theta_j$: 5$^\circ$, 10$^\circ$, and 20$^\circ$ (since this not-well-known parameter has the greatest effect in determining the determining the detectability of afterglows).

For $\theta_j=5^\circ$ (in which case 0.095\% of GRBs will occur on-axis), we found that 226 afterglows (2.3\%) would be detectable (assuming the field they are located in is observed;) and that 43\% of the detectable afterglows (AGs) would never exceeded rank 3 (i.e., they were classified as rank 3 in both phases 1 and 2).
Among the 226, 120 AGs were detectable in both phases, 47 AGs in only phase 1, and 59 AGs in only phase 2.
For $\theta_j=10^\circ$ (in which case 0.38\% of GRBs will occur on-axis), we found that 583 AGs (5.8\%) would be detectable, and that 40\% would never exceeded rank 3.
Among the 583, 398 AGs were detectable in both phases, 77 AGs in only phase 1, and 108 in only phase 2.
For $\theta_j=20^\circ$ (in which case 1.5\% of GRBs will occur on-axis), we found that 1399 afterglows (14\%) would be detectable, and that 30\% would never exceed rank 3.
Among the 1399, 1030 AGs were detectable in both phases, 201 in only phase 1, and 168 in only phase 2.
We also found that very few of the detectable AGs will exhibit fading with a significance below $1\sigma$; for $\theta_j=2.5^\circ,5^\circ,10^\circ$, the rates were 4.2\%, 1.1\%, and 0.3\%.
Although 16\% of constant sources should be expected to exhibit fading with a significance $\sigma\geq1$ purely by chance (which is in agreement with the observed number of ``fading'' rank 3 sources: 8/51 = 16\%), real AGs will almost always exhibit fading at this significance or higher across timescales corresponding to the spacing of our phase 1 and phase 2 observations.
We will use this finding to prioritize phase 3 follow-up of sources with these characteristics.

\subsection{Changes for O3}

The lessons from O2 do not necessitate any major changes from the XRT perspective.
To date, we have focused \swift\ follow-up on the XRT, but GW 170817 has shown that UVOT is a crucial discovery instrument (see \citealt{Evans2017}).
However, a challenge can arise from the UVOT field of view being smaller than that of the XRT.
To address this, for CBC triggers in O3, we have modified our target selection criteria.
Now, fields are initially selected based on tiling the XRT (as it is not practical to tile the smaller UVOT), but any field which was selected because it contained a potential host galaxy can now be offset or split into multiple fields. 
This will ensure that the galaxy or galaxies in question fall entirely within the UVOT field of view.
We have also made changes to the way in which selected fields are organized into an observing plan in order to achieve greater efficiency (in that less time is spent slewing) in the coverage of the LVC region.

As another minor change, the exposure time of phase 1 observations has been increased from 60 to 80 s.
Post-processing of the initial 60 s observations occasionally reduces their effective exposes to less than 60 s (e.g., the removal of periods of high background), and/or exposures can also be shortened due to uncertainties in the estimated durations of slews.
The extra allocated 20 s will guarantee that all initial exposures reach effective times of at least 60 s.

In regards to the evaluation of our source ranking criteria: although we will not be changing  source ranking criteria, we will begin to include the significance of any fading behavior in detected sources in GCN notices.
This will allow us to differentiate ``interesting'' rank 3 sources from the ``uninteresting'' ones, without lowering the threshold criteria for rank 2 sources.
We would like to remind the astronomical community that real afterglows (at typical distances) have an estimated 30-40\% of being assigned rank 3, and that in almost all cases we should be able to detect fading at at least 1$\sigma$ significance.
Therefore, fading rank 3 sources should not be neglected in potential multiwavelength follow-up, especially in the absence of higher-rank targets.

Lastly, in future searches we will be able to make use of an additional tool to detect an afterglow: the \swift\ Gravitational Wave Galaxy Survey (SGWGS).
For this campaign, we selected the 20,000 most luminous galaxies in the GWGC (i.e., within 100 Mpc) and are seeking to observe each for at least 1 ks. 
The 20,000 galaxies chosen equate to roughly 50\% of the total luminosity in the GWGC.
This gives us beneficial pre-imaging information which will allow us to determine the number of pre-existing sources (which will make the identification of transients more reliable in those fields), and also will give us a set of reference images for calculating upper limits or performing UVOT difference imaging.
This will also drastically reduce the number of uncatalogued X-ray and UV/optical sources which a potential afterglow might be mis-categorized as. 
Currently, about 50\% of the selected galaxies have been imaged.
The details of the SGWGS will be discussed in a later publication (Tohuvavohu et al.\ in prep).

\section{SUMMARY}
We reported on the follow-up searches carried out by the Neil Gehrels \swift\ Observatory for GW triggers in the second LVC observing run (O2) in 2017, and described the search strategy and source analysis methods employed.
The \swift-XRT observed 3998 fields containing potential host galaxies to the GW events, covering a total of 449 deg$^2$.
We described the properties of the 157 (non-counterpart) X-ray sources detected, and the detection rates of sources of each rank.
We found that 17\% of the sources exhibited variability, with the majority (7/10) of identified variable sources being AGNs.
Thus, we expect the rate of unrelated transients to be manageably low.
We re-examined the \swift\ follow-up strategy and our source ranking criteria and discuss minor improvements to be implemented.

\facility{the Neil Gehrels {\sl Swift} Observatory}
\software{HEAsoft (v6.22; \citealt{HEASARC2014}), afterglowpy (v0.6.4; \citealt{Ryan2019})}

\acknowledgements
NJK would like to acknowledge support from NASA Grant 80NSSC19K0408.
PAE, APB, JPO, and KLP acknowledge support from the UK Space Agency.
SRO gratefully acknowledges the support of the Leverhulme Trust Early Career Fellowship. 
The Dark Cosmology Centre was funded by the Danish National Research Foundation. DBM is supported by research grant 19054 from Villum Fonden.
AD acknowledges financial contribution from the agreement ASI-INAF n.2017-14-H.0.
The authors would also like to thank the anonymous referee for their useful suggestions which helped to improve the paper.

\bibliographystyle{aasjournal}
\bibliography{swifto2}

\input{catalog.tex}

\appendix

\section{Details of LVC Triggers and \swift\ Follow-up in O2}

\subsection{G268556 / LVC 170104}
A CBC trigger occurred at 10:11:59 UT, with an estimated distance $d=737\pm201$ Mpc.
The first pre-planned science target (PPST) list was uploaded at 21:37 UT, and the first \swift\ observation occurred at T0 + 50.32 ks.
Of the 600 planned tiles, 293 were observed, covering 4.7\% of the raw (and 4.6\% of the convolved) LVC probability region.
30 X-ray sources were detected: 6 known (rank 4) sources and 24 unlikely afterglows (rank 3 sources).

ATLAS identified a possible counterpart (ATLAS17aeu) which \swift\ followed-up and found to be fading in X-rays (GCNs 20390, 20415).
UVOT found nothing above 3$\sigma$ at the source's position, but did note a 2.9$\sigma$ source in the W2 filter which appeared to be fading (GCN 20400).

The BAT team performed a coincidence search but found nothing (GCN 20422).

\swift\ followed-up another possible counterpart, iPTF17cw, but found no X-rays in 1.3 ks worth of observations (GCN 20473).

7 GCN circulars were sent by the \swift\ team in relation to this trigger: 20473, 20422, 20415, 20400, 20390, 20371.

\subsection{G270580 / LVC 170120}
LVC detected an unmodeled burst trigger at 12:31:00 UT.
The first PPST list was uploaded at 17:35 UT, and the first \swift\ observation occurred at T0 + 19.9 ks.
Of the 159 planned tiles, 136 were observed, covering 1.1\% of the raw (and 14\% of the convolved) LVC probability region.
3 X-ray sources were detected: 2 known (rank 4) source and 1 unlikely afterglow (rank 3 source).

4 GCNs circulars were sent by the \swift\ team.  
The highlights are as follows:
GCN 20494: 18\% of the LVC error region was covered by the BAT at the time of trigger, but no significant detection was found.
GCN 20495: The above-mentioned XRT sources were reported.
GCN 20523: A report on the XRT follow-up of possible counterpart PS17yt; no source was detected in 3 ks of observation.
GCN 20536: A report on the UVOT follow-up of the same source, which was found at  magnitude 20.9.

\subsection{G274296 / LVC 170217}
LVC detected an unmodeled burst trigger at 06:05:53 UT.
No tiling was performed, since the trigger did not meet our trigger criteria.
The BAT covered 65.5\% of the LVC error region at the time of trigger, but no significant detections were found (GCN 20744).

\subsection{G275404 / LVC 170225}
LVC detected a CBC trigger at 18:30:21 UT, with an estimated distance $d=412\pm169$ Mpc.
It was reported as having a probability of being EM-bright of 0.9, and a $P_{\rm NS}=1$.
The first PPST list was uploaded at 23:08 UT, and the first \swift\ observation occurred at T0 + 17.3 ks.
Of the 1402 planned tiles, 117 were observed, covering 1.8\% of the raw (and 4\% of the convolved) LVC probability region.
3 X-ray sources were detected: 1 known (rank 4) source and 2 unlikely afterglows (rank 3 sources).
The trigger was revised on 02/26--00:20, and again at 02/26--06:16.
Phase 1 was carried out, but trigger G275697 occurred when phase 2 was due to start, so we decided to prioritize the more recent and nearby event (described in following section).

3 GCN circulars were sent by the \swift\ team.
GCN 20746: The BAT covered 18.68\% of the LVC probability region at the time of trigger, but no significant detections were found.
GCN 20752: A report on the above-mentioned XRT sources.
GCN 20842: A report on the XRT follow-up of the possible counterparts coincident with the position of the transient reported by AGILE, AGL J1914+1043:  GL 191032+075314, and GRS 1915+105.

\subsection{G275697 / LVC 170227}
LVC detected a CBC trigger at 18:57:31 UT, with an estimated distance $d=193\pm61$ Mpc.
It was reported as having a probability of being EM-bright of 1.
Of the 1414 planned tiles, 1408 tiles were observed.
This is the first trigger for which phase 1 was carried out in full (e.g., without a reduction to the number of tiles), and the first (and only) trigger for which the full 3-phase observing plan was carried out.
No negative effects on the spacecraft (e.g., due to the high rate of slewing) were recorded.
The trigger was subsequently retracted, after our follow-up search was performed.

8 GCN circulars were sent by the \swift\ team.
GCN 20772: The BAT covered 8\% of the LVC probability region at the time of trigger, but no significant detections were found.
GCNs 20773, 20798, 20807, 20812, 20821 were reports on sources found by the XRT.
GCN 20841 and 20884: The XRT team reported on the observed fading and subsequent cessation of fading of a source.

\subsection{G277583 / LVC 170313}
LVC detected an unmodeled burst trigger at 22:40:09 UT.
No tiling was performed, since the trigger did not meet our trigger criteria.
The BAT covered 19.2\% of the LVC probability region at trigger time, but no significant detections were found (GCN 20883).

\subsection{G284239 / LVC 170502}
LVC detected an unmodeled burst trigger at 22:26:07 UT.
No tiling was performed, since the trigger did not meet our trigger criteria.
0\% of the LVC probability region was covered by the BAT at the time of trigger, and no significant detections were found (GCN 21065).

\subsection{G288732 / LVC 170608}
LVC detected a CBC trigger at 02:01:16 UT, with an estimated distance $d=320\pm98$ Mpc.
No tiling was initially performed, since the trigger did not meet our trigger criteria.
However, a 4-point tiling was performed at T0 + 62.8 ks to follow-up a possible counterpart detected by the \fermi-LAT.
7 X-ray sources were detected: 2 known (rank 4) sources and 5 unlikely afterglows (rank 3 sources).

3 GCN circulars were sent by the \swift\ team.
GCN 21233: described the XRT sources detected in the follow-up of the LAT source.
GCN 21234: no significant detections were found in the BAT data (the BAT covered 0\% of the LVC probability region at the time of trigger).
GCN 21235: summarized the UVOT follow-up; no compelling sources were found.

The following day the region of sky entered \swift's Sun observing constraint (until September), so no further follow-up was conducted.

\subsection{G296853 / LVC 170809}
LVC detected a CBC trigger at 08:28:21 UT, with an estimated distance $d=1086\pm302$ Mpc.
No tiling was performed, since the trigger did not meet our trigger criteria.
0\% of the LVC probability region was covered by the BAT at the time of trigger (GCN 21436).

\subsection{G297595 / LVC 170814}
LVC detected a CBC trigger at 10:30:43 UT, with an estimated distance $d=534\pm131$ Mpc.
The first PPST list was uploaded at 18:22 UT, and the first observation occurred at T0 + 31.2 ks.
Of the 726 planned tiles, 643 were observed, covering 24\% of the raw (and 36\% of the convolved) LVC probability region.
41 X-ray sources were detected: 26 known (rank 4) sources and 15 unlikely afterglows (rank 3 sources).

2 GCNs were sent by the \swift\ team.
GCN 21483: 0\% of the LVC probability region was covered by the BAT at the time of trigger; a 5.4$\sigma$ spike was detected at T0+27 seconds, but it is likely noise.
GCN 21503: reported on the XRT sources.

\subsection{G298048 / LVC 170817}
LVC detected a CBC trigger at 12:41:04 UT, with an estimated distance $d=39\pm7$ Mpc.
The first PPST list was promptly uploaded and the first observation occurred at T0 + 3.3 ks.
Of the 2966 tilings planned, 744 were observed, covering 2.5\% of the raw (and 94\% of the convolved) LVC probability region.
This was the epochal GW 170817, the first GW event for which the EM counterpart was discovered and identified.
\swift\ and \nustar\ observations of this event are discussed in-depth by \citet{Evans2017}.

\subsection{G298389 / LVC 170819}
LVC detected an unmodeled burst trigger at 15:50:46 UT.
No tiling was performed, as we decided to prioritize the follow-up search for the previous trigger (GW 170817).
0\% of the LVC probability region was covered by the BAT at the time of trigger, and no significant detections were found (GCN 21622).

\subsection{G298936 / LVC 170823}
LVC detected a CBC trigger at 13:13:58 UT, with an estimated distance $d=1738\pm477$ Mpc.
No tiling was performed, as the source was too far away for galaxy targeting, the event was a BBH, and we were still following-up with GW 170817.
9.5\% of the LVC probability region was covered by the BAT at the time of trigger; a 5.1$\sigma$ spike was detected at T0-30 s, but it was probably not astrophysical in origin (GCN 21665).

\subsection{G299232 / LVC 170825}
LVC detected a CBC trigger at 13:13:37 UT, with an estimated distance $d=339\pm109$ Mpc.
The first PPST list was uploaded at 15:32 UT, and the first \swift\ observation was carried out at T0 + 11.1 ks.
Of the 1096 tilings planned, 653 were observed, covering 8.3\% of the raw (and 16.0\% of the convolved) LVC probability region.
The list of tiles for this plan had to be reduced due to XRT temperature considerations.
51 X-ray sources were detected: 1 interesting (rank 2) source, 30 unlikely afterglows (rank 3 sources), and 20 known (rank 4) sources.
The rank 2 source was 1RXS J014709.9+234529 (an RS CVn variable star).

4 GCN circulars were sent by the \swift\ team.
GCN 21704: 11.56\% of the LVC probability region was covered by BAT at the time of trigger, but no significant detections were found.
GCN 21733: the optical transient \swift\ J014008.5+343403.6 was discovered with the UVOT at magnitude 18, but with no X-ray counterpart.
GCN 21758: the UVOT counterpart was observed again, but was not detected.
GCN 21844: reported on the XRT sources.

\end{document}

%% file: catalog.tex
\startlongtable
\begin{deluxetable*}{ccccccccccc}
\tabletypesize{\scriptsize}
\tablecolumns{12}
\tablecaption{\label{tab:cattable}Catalog of X-ray sources detected in the follow-up searches for O2 GW triggers.}
\tablehead{
\colhead{GW } & 
\colhead{Rank} &
\colhead{RA/Dec} & 
\colhead{Err} & 
\colhead{Peak Rate}  & 
\colhead{Peak Flux} & 
\colhead{Simbad} & 
\colhead{Known} & 
\colhead{Near} &
\colhead{Near} &
\colhead{Fading}
\\
\colhead{\#} & 
\colhead{} &
\colhead{(J2000)} & 
\colhead{($''$)} & 
\colhead{(cts s$^{-1}$)} & 
\colhead{(erg cm$^{-2}$ s$^{-1}$)} & 
\colhead{src type} & 
\colhead{X} & 
\colhead{gal.} &
\colhead{2MASS} &
\colhead{($\sigma$)}
}
\startdata
8 & 4 & $00^h 00^m 31\fs64$ +$68^\circ 15' 00\farcs4$ & 4.9 & 0.028 ($\pm$0.009) & 1.2 ($\pm$0.4) $\times10^{-12}$ & Star & Y & 2 & 1 & 0 \\
8 & 4 & $01^h 09^m 44\fs03$ +$73^\circ 11' 58\farcs5$ & 5.6 & 0.14 ($\pm$0.05) & 6.1 ($\pm$2.2) $\times10^{-12}$ & Seyfert 1 & Y & 1 & 1 & 0 \\
17 & 3 & $01^h 20^m 16\fs99$ +$12^\circ 03' 20\farcs3$ & 6.1 & 0.013 ($\pm$0.005) & 5.7 ($\pm$2.1) $\times10^{-13}$ & Radio & N & 0 & 1 & 0 \\
17 & 3 & $01^h 24^m 24\fs94$ +$08^\circ 24' 03\farcs1$ & 5.2 & 0.033 ($\pm$0.010) & 1.4 ($\pm$0.4) $\times10^{-12}$ & EB*WUMa & N & 1 & 1 & 0 \\
17 & 4 & $01^h 24^m 42\fs22$ +$08^\circ 51' 24\farcs6$ & 4.5 & 0.046 ($\pm$0.012) & 2.0 ($\pm$0.5) $\times10^{-12}$ & Seyfert & Y & 1 & 1 & 0 \\
17 & 4 & $01^h 39^m 10\fs99$ +$34^\circ 33' 40\farcs3$ & 6.6 & 0.068 ($\pm$0.028) & 2.9 ($\pm$1.2) $\times10^{-12}$ & X & Y & 0 & 1 & 0 \\
17 & 3 & $01^h 40^m 17\fs34$ +$23^\circ 27' 21\farcs3$ & 6.4 & 0.020 ($\pm$0.008) & 9 ($\pm$3) $\times10^{-13}$ & QSO & N & 0 & 1 & 0 \\
17 & 4 & $01^h 44^m 47\fs20$ +$32^\circ 32' 56\farcs3$ & 4.9 & 0.07 ($\pm$0.04) & 3.0 ($\pm$1.8) $\times10^{-12}$ & Unknown & Y & 0 & 1 & 0 \\
17 & 3 & $01^h 45^m 42\fs52$ +$32^\circ 43' 39\farcs8$ & 6.8 & 0.028 ($\pm$0.011) & 1.2 ($\pm$0.5) $\times10^{-12}$ & Unknown & N & 0 & 1 & 0 \\
17 & 4 & $01^h 46^m 33\fs86$ +$33^\circ 17' 08\farcs9$ & 6.3 & 0.12 ($\pm$0.05) & 5.2 ($\pm$2.2) $\times10^{-12}$ & BYDra & Y & 0 & 3 & 0 \\
17 & 2 & $01^h 47^m 09\fs95$ +$23^\circ 45' 30\farcs5$ & 4.1 & 0.076 ($\pm$0.015) & 3.3 ($\pm$0.6) $\times10^{-12}$ & RSCVn & Y & 0 & 1 & 0 \\
17 & 3 & $01^h 55^m 19\fs94$ +$41^\circ 27' 00\farcs3$ & 6.1 & 0.020 ($\pm$0.012) & 9 ($\pm$5) $\times10^{-13}$ & Unknown & N & 1 & 0 & 0 \\
17 & 4 & $01^h 55^m 35\fs91$ +$31^\circ 15' 15\farcs2$ & 5.2 & 0.036 ($\pm$0.010) & 1.5 ($\pm$0.4) $\times10^{-12}$ & Seyfert 1 & Y & 2 & 1 & 0 \\
17 & 4 & $01^h 57^m 15\fs37$ +$31^\circ 54' 14\farcs6$ & 5.7 & 0.027 ($\pm$0.010) & 1.1 ($\pm$0.4) $\times10^{-12}$ & Radio & Y & 0 & 1 & 0 \\
17 & 4 & $01^h 57^m 56\fs31$ +$40^\circ 24' 17\farcs6$ & 6.8 & 0.022 ($\pm$0.009) & 10 ($\pm$4) $\times10^{-13}$ & Unknown & Y & 1 & 1 & 0 \\
17 & 3 & $01^h 58^m 48\fs10$ +$36^\circ 21' 39\farcs6$ & 5.1 & 0.024 ($\pm$0.009) & 1.0 ($\pm$0.4) $\times10^{-12}$ & Unknown & N & 3 & 1 & 0 \\
17 & 3 & $02^h 00^m 38\fs66$ +$44^\circ 27' 17\farcs8$ & 5.7 & 0.013 ($\pm$0.006) & 5.6 ($\pm$2.6) $\times10^{-13}$ & QSO & N & 0 & 0 & 0 \\
17 & 4 & $02^h 01^m 06\fs90$ +$44^\circ 08' 42\farcs1$ & 4.7 & 0.062 ($\pm$0.014) & 2.7 ($\pm$0.6) $\times10^{-12}$ & Unknown & Y & 0 & 1 & 0 \\
17 & 3 & $02^h 01^m 09\fs81$ +$44^\circ 10' 16\farcs2$ & 5.8 & 0.018 ($\pm$0.008) & 8 ($\pm$3) $\times10^{-13}$ & (AGN; MQ) & N & 0 & 1 & 0 \\
17 & 3 & $02^h 02^m 01\fs50$ +$39^\circ 43' 19\farcs5$ & 7.5 & 0.017 ($\pm$0.007) & 7 ($\pm$3) $\times10^{-13}$ & (AGN; MQ) & N & 1 & 1 & 0 \\
17 & 3 & $02^h 03^m 24\fs54$ +$39^\circ 51' 19\farcs7$ & 7.9 & 0.04 ($\pm$0.03) & 1.8 ($\pm$1.3) $\times10^{-12}$ & Unknown & N & 0 & 1 & 1.2 \\
17 & 4 & $02^h 12^m 26\fs58$ +$52^\circ 09' 51\farcs9$ & 5.4 & 0.047 ($\pm$0.013) & 2.0 ($\pm$0.6) $\times10^{-12}$ & AGN Candidate & Y & 0 & 1 & 0 \\
17 & 4 & $02^h 14^m 17\fs97$ +$51^\circ 44' 42\farcs9$ & 4.9 & 0.29 ($\pm$0.07) & 1.3 ($\pm$0.3) $\times10^{-11}$ & Unknown & Y & 5 & 0 & 0 \\
17 & 4 & $02^h 18^m 05\fs26$ +$39^\circ 17' 44\farcs4$ & 7.9 & 0.045 ($\pm$0.012) & 2.0 ($\pm$0.5) $\times10^{-12}$ & X & Y & 4 & 1 & 0 \\
13 & 4 & $02^h 18^m 25\fs40$ --$50^\circ 13' 29\farcs0$ & 5.7 & 0.017 ($\pm$0.006) & 7.5 ($\pm$2.7) $\times10^{-13}$ & X (AGN; MQ) & Y & 3 & 0 & 0 \\
13 & 4 & $02^h 18^m 30\fs66$ --$48^\circ 06' 53\farcs1$ & 4.9 & 0.08 ($\pm$0.05) & 3.3 ($\pm$2.0) $\times10^{-12}$ & X & Y & 0 & 1 & 1.2 \\
13 & 3 & $02^h 19^m 28\fs48$ --$48^\circ 31' 15\farcs2$ & 5.6 & 0.017 ($\pm$0.007) & 7 ($\pm$3) $\times10^{-13}$ & Galaxy & N & 2 & 1 & 0 \\
17 & 4 & $02^h 19^m 52\fs91$ +$43^\circ 55' 18\farcs3$ & 6.8 & 0.018 ($\pm$0.010) & 8 ($\pm$4) $\times10^{-13}$ & Unknown & Y & 1 & 1 & 0 \\
17 & 3 & $02^h 20^m 14\fs55$ +$50^\circ 44' 44\farcs1$ & 9.8 & 0.020 ($\pm$0.008) & 9 ($\pm$3) $\times10^{-13}$ & Unknown & N & 0 & 2 & 0 \\
13 & 3 & $02^h 20^m 23\fs62$ --$51^\circ 24' 00\farcs8$ & 5.9 & 0.019 ($\pm$0.008) & 8 ($\pm$3) $\times10^{-13}$ & Unknown & N & 0 & 0 & 0 \\
17 & 4 & $02^h 20^m 35\fs63$ +$50^\circ 44' 11\farcs6$ & 5.2 & 0.017 ($\pm$0.007) & 7 ($\pm$3) $\times10^{-13}$ & Unknown & Y & 0 & 1 & 0 \\
17 & 3 & $02^h 21^m 26\fs96$ +$51^\circ 26' 11\farcs9$ & 5.1 & 0.013 ($\pm$0.005) & 5.4 ($\pm$2.0) $\times10^{-13}$ & Unknown & N & 0 & 1 & 0 \\
17 & 4 & $02^h 22^m 38\fs72$ +$43^\circ 02' 09\farcs6$ & 4.3 & 0.14 ($\pm$0.06) & 5.8 ($\pm$2.4) $\times10^{-12}$ & Unknown & Y & 6 & 0 & 1.4 \\
17 & 3 & $02^h 23^m 05\fs28$ +$43^\circ 30' 48\farcs4$ & 6.1 & 0.022 ($\pm$0.007) & 9 ($\pm$3) $\times10^{-13}$ & Unknown & N & 0 & 1 & 0 \\
13 & 3 & $02^h 23^m 29\fs89$ --$50^\circ 29' 31\farcs7$ & 5.8 & .01 ($\pm$.004) & 4.1 ($\pm$1.8) $\times10^{-13}$ & Unknown & N & 0 & 0 & 0 \\
13 & 3 & $02^h 24^m 11\fs57$ --$49^\circ 53' 04\farcs6$ & 5.1 & 0.035 ($\pm$0.013) & 1.5 ($\pm$0.5) $\times10^{-12}$ & Unknown & N & 0 & 1 & 0 \\
13 & 3 & $02^h 25^m 02\fs96$ --$53^\circ 52' 59\farcs6$ & 5.1 & 0.045 ($\pm$0.011) & 1.9 ($\pm$0.5) $\times10^{-12}$ & Unknown & N & 1 & 1 & 0 \\
13 & 3 & $02^h 25^m 53\fs44$ --$48^\circ 26' 28\farcs9$ & 6.1 & 0.022 ($\pm$0.009) & 9 ($\pm$4) $\times10^{-13}$ & Unknown & N & 0 & 1 & 0 \\
13 & 4 & $02^h 26^m 46\fs93$ --$50^\circ 37' 56\farcs6$ & 6 & 0.014 ($\pm$0.007) & 6.0 ($\pm$2.8) $\times10^{-13}$ & X & Y & 0 & 0 & 0 \\
13 & 3 & $02^h 27^m 03\fs80$ --$49^\circ 48' 08\farcs1$ & 6 & 0.022 ($\pm$0.008) & 9 ($\pm$3) $\times10^{-13}$ & Unknown & N & 0 & 0 & 0 \\
13 & 3 & $02^h 28^m 52\fs21$ --$49^\circ 21' 36\farcs8$ & 6 & 0.05 ($\pm$0.04) & 2.0 ($\pm$1.5) $\times10^{-12}$ & (AGN; MQ) & N & 0 & 1 & 0 \\
13 & 4 & $02^h 30^m 00\fs37$ --$54^\circ 01' 19\farcs8$ & 7.8 & 0.046 ($\pm$0.016) & 2.0 ($\pm$0.7) $\times10^{-12}$ & Unknown & Y & 0 & 0 & 0 \\
13 & 4 & $02^h 30^m 20\fs85$ --$54^\circ 15' 08\farcs8$ & 5.2 & 0.034 ($\pm$0.012) & 1.5 ($\pm$0.5) $\times10^{-12}$ & Unknown & Y & 0 & 1 & 0 \\
13 & 4 & $02^h 31^m 31\fs13$ --$48^\circ 26' 57\farcs4$ & 5 & 0.031 ($\pm$0.011) & 1.3 ($\pm$0.5) $\times10^{-12}$ & AGN Candidate & Y & 0 & 1 & 0 \\
13 & 4 & $02^h 34^m 31\fs24$ --$46^\circ 31' 59\farcs0$ & 6.3 & 0.11 ($\pm$0.05) & 4.9 ($\pm$2.3) $\times10^{-12}$ & Galaxy & Y & 0 & 1 & 0 \\
13 & 3 & $02^h 35^m 29\fs57$ --$50^\circ 21' 21\farcs2$ & 6.3 & 0.05 ($\pm$0.04) & 2.3 ($\pm$1.7) $\times10^{-12}$ & Unknown & N & 1 & 2 & 1.4 \\
13 & 4 & $02^h 38^m 19\fs17$ --$52^\circ 11' 33\farcs7$ & 3.6 & 1.06 ($\pm$0.07) & 4.5 ($\pm$0.3) $\times10^{-11}$ & Seyfert 1 & Y & 3 & 1 & 0 \\
13 & 4 & $02^h 38^m 20\fs92$ --$53^\circ 25' 32\farcs5$ & 5.1 & 0.029 ($\pm$0.010) & 1.2 ($\pm$0.4) $\times10^{-12}$ & AGN Candidate & Y & 1 & 1 & 0 \\
13 & 4 & $02^h 42^m 12\fs18$ --$54^\circ 57' 13\farcs5$ & 5.7 & 0.022 ($\pm$0.008) & 9 ($\pm$4) $\times10^{-13}$ & Galaxy & Y & 0 & 1 & 0 \\
13 & 4 & $02^h 42^m 36\fs78$ --$55^\circ 06' 35\farcs7$ & 4.8 & 0.035 ($\pm$0.010) & 1.5 ($\pm$0.4) $\times10^{-12}$ & Galaxy & Y & 2 & 1 & 0 \\
13 & 4 & $02^h 45^m 12\fs97$ --$46^\circ 27' 55\farcs3$ & 6.7 & 0.017 ($\pm$0.008) & 7 ($\pm$3) $\times10^{-13}$ & Galaxy & Y & 5 & 2 & 0 \\
13 & 4 & $02^h 45^m 53\fs78$ --$44^\circ 59' 38\farcs2$ & 5.4 & 0.08 ($\pm$0.05) & 3.5 ($\pm$2.1) $\times10^{-12}$ & QSO & Y & 0 & 1 & 0 \\
13 & 4 & $02^h 51^m 11\fs70$ --$47^\circ 53' 14\farcs1$ & 5 & 0.047 ($\pm$0.011) & 2.0 ($\pm$0.5) $\times10^{-12}$ & RSCVn & Y & 0 & 1 & 0 \\
13 & 3 & $03^h 09^m 51\fs89$ --$43^\circ 31' 15\farcs6$ & 6.8 & 0.011 ($\pm$0.005) & 4.8 ($\pm$2.1) $\times10^{-13}$ & (AGN; MQ) & N & 0 & 1 & 0 \\
13 & 4 & $03^h 12^m 25\fs39$ --$44^\circ 25' 17\farcs3$ & 5.1 & 0.21 ($\pm$0.03) & 9.0 ($\pm$1.4) $\times10^{-12}$ & PM* & Y & 0 & 1 & 0 \\
13 & 3 & $03^h 13^m 41\fs96$ --$44^\circ 41' 08\farcs8$ & 6 & 0.020 ($\pm$0.008) & 9 ($\pm$3) $\times10^{-13}$ & Unknown & N & 0 & 0 & 0 \\
13 & 4 & $03^h 13^m 42\fs35$ --$41^\circ 59' 39\farcs5$ & 4.9 & 0.035 ($\pm$0.011) & 1.5 ($\pm$0.5) $\times10^{-12}$ & AGN Candidate & Y & 1 & 1 & 0 \\
13 & 4 & $03^h 14^m 51\fs05$ --$42^\circ 02' 52\farcs7$ & 4.4 & 0.036 ($\pm$0.011) & 1.6 ($\pm$0.5) $\times10^{-12}$ & (AGN; MQ) & Y & 0 & 0 & 0 \\
13 & 4 & $03^h 14^m 55\fs60$ --$42^\circ 41' 00\farcs8$ & 5 & 0.071 ($\pm$0.017) & 3.0 ($\pm$0.7) $\times10^{-12}$ & Seyfert 1 & Y & 2 & 1 & 0 \\
13 & 4 & $03^h 16^m 45\fs04$ --$42^\circ 31' 30\farcs7$ & 5.7 & 0.050 ($\pm$0.016) & 2.1 ($\pm$0.7) $\times10^{-12}$ & PM* & Y & 0 & 1 & 0 \\
13 & 3 & $03^h 17^m 53\fs85$ --$44^\circ 12' 03\farcs9$ & 6.7 & 0.027 ($\pm$0.009) & 1.2 ($\pm$0.4) $\times10^{-12}$ & Unknown & N & 1 & 0 & 0 \\
13 & 3 & $03^h 17^m 56\fs06$ --$39^\circ 05' 38\farcs2$ & 5 & 0.034 ($\pm$0.012) & 1.5 ($\pm$0.5) $\times10^{-12}$ & Unknown & N & 0 & 1 & 0 \\
13 & 4 & $03^h 17^m 56\fs75$ --$44^\circ 13' 46\farcs7$ & 8.3 & 0.270 ($\pm$0.026) & 1.2 ($\pm$0.11) $\times10^{-11}$ & Unknown & Y & 2 & 2 & 0 \\
13 & 4 & $03^h 17^m 57\fs19$ --$44^\circ 15' 39\farcs3$ & 5.8 & 0.176 ($\pm$0.023) & 7.5 ($\pm$1.0) $\times10^{-12}$ & QSO & Y & 2 & 2 & 0 \\
13 & 4 & $03^h 17^m 57\fs53$ --$44^\circ 14' 19\farcs7$ & 5.2 & 0.180 ($\pm$0.023) & 7.7 ($\pm$1.0) $\times10^{-12}$ & QSO & Y & 2 & 2 & 0 \\
8 & 4 & $03^h 17^m 57\fs54$ --$44^\circ 14' 15\farcs8$ & 4.7 & 0.12 ($\pm$0.06) & 5.0 ($\pm$2.6) $\times10^{-12}$ & EB*WUMa & Y & 0 & 1 & 1.3 \\
13 & 3 & $03^h 18^m 00\fs85$ --$44^\circ 12' 06\farcs8$ & 8.4 & 0.037 ($\pm$0.011) & 1.6 ($\pm$0.5) $\times10^{-12}$ & Unknown & N & 1 & 0 & 0 \\
13 & 4 & $03^h 18^m 01\fs64$ --$44^\circ 13' 43\farcs2$ & 8.5 & 0.041 ($\pm$0.011) & 1.8 ($\pm$0.5) $\times10^{-12}$ & Unknown & Y & 3 & 1 & 0 \\
13 & 4 & $03^h 18^m 01\fs70$ --$44^\circ 14' 40\farcs1$ & 9.6 & 0.070 ($\pm$0.014) & 3.0 ($\pm$0.6) $\times10^{-12}$ & Unknown & Y & 4 & 0 & 0 \\
13 & 3 & $03^h 18^m 09\fs58$ --$44^\circ 11' 44\farcs5$ & 6.2 & 0.016 ($\pm$0.007) & 7 ($\pm$3) $\times10^{-13}$ & Unknown & N & 0 & 0 & 0 \\
13 & 4 & $03^h 32^m 49\fs09$ --$26^\circ 02' 45\farcs0$ & 6.8 & 0.021 ($\pm$0.008) & 9 ($\pm$3) $\times10^{-13}$ & QSO & Y & 0 & 0 & 0 \\
17 & 3 & $03^h 49^m 43\fs13$ +$75^\circ 16' 07\farcs6$ & 5.9 & 0.017 ($\pm$0.007) & 7 ($\pm$3) $\times10^{-13}$ & Unknown & N & 2 & 1 & 0 \\
17 & 4 & $04^h 57^m 53\fs20$ +$80^\circ 06' 50\farcs9$ & 8.4 & 0.021 ($\pm$0.008) & 9 ($\pm$3) $\times10^{-13}$ & Unknown & Y & 1 & 1 & 0 \\
5 & 4 & $07^h 20^m 28\fs70$ +$71^\circ 32' 35\farcs0$ & 5.7 & 0.18 ($\pm$0.07) & 8 ($\pm$3) $\times10^{-12}$ & Unknown & Y & 2 & 0 & 0 \\
4 & 4 & $08^h 52^m 20\fs14$ +$47^\circ 34' 57\farcs8$ & 5.4 & 0.18 ($\pm$0.07) & 7.9 ($\pm$2.9) $\times10^{-12}$ & QSO & Y & 0 & 0 & 0 \\
4 & 4 & $08^h 53^m 46\fs02$ +$47^\circ 18' 42\farcs0$ & 5 & 0.21 ($\pm$0.07) & 9.1 ($\pm$2.9) $\times10^{-12}$ & RotV* & Y & 0 & 1 & 0 \\
8 & 2 & $09^h 29^m 46\fs62$ +$02^\circ 03' 47\farcs8$ & 6 & 0.05 ($\pm$0.04) & 2.3 ($\pm$1.8) $\times10^{-12}$ & LINER & N & 5 & 1 & 4.1 \\
8 & 4 & $09^h 47^m 02\fs28$ --$05^\circ 56' 50\farcs6$ & 5.7 & 0.017 ($\pm$0.009) & 7 ($\pm$4) $\times10^{-13}$ & Galaxy & Y & 1 & 1 & 0 \\
8 & 4 & $09^h 57^m 17\fs97$ --$13^\circ 50' 00\farcs2$ & 6.1 & 0.025 ($\pm$0.015) & 1.1 ($\pm$0.6) $\times10^{-12}$ & QSO & Y & 2 & 0 & 0 \\
8 & 3 & $09^h 58^m 13\fs62$ --$05^\circ 24' 30\farcs6$ & 6 & 0.019 ($\pm$0.008) & 8 ($\pm$3) $\times10^{-13}$ & Galaxy & N & 0 & 1 & 0 \\
8 & 4 & $09^h 58^m 33\fs58$ --$05^\circ 21' 37\farcs9$ & 4.7 & 0.07 ($\pm$0.04) & 3.0 ($\pm$1.8) $\times10^{-12}$ & Unknown & Y & 0 & 1 & 0 \\
8 & 3 & $10^h 02^m 45\fs70$ --$16^\circ 11' 48\farcs5$ & 5.6 & 0.020 ($\pm$0.008) & 9 ($\pm$4) $\times10^{-13}$ & (AGN; MQ) & N & 0 & 0 & 0 \\
8 & 4 & $10^h 03^m 41\fs72$ --$15^\circ 08' 01\farcs3$ & 5.6 & 0.07 ($\pm$0.05) & 3.1 ($\pm$2.4) $\times10^{-12}$ & QSO & Y & 2 & 1 & 1.1 \\
8 & 4 & $10^h 08^m 03\fs01$ --$14^\circ 59' 00\farcs9$ & 4.6 & 0.07 ($\pm$0.04) & 3.0 ($\pm$1.8) $\times10^{-12}$ & Seyfert 1 & Y & 1 & 2 & 0 \\
8 & 4 & $10^h 11^m 59\fs30$ --$16^\circ 36' 31\farcs6$ & 5 & 0.039 ($\pm$0.011) & 1.7 ($\pm$0.5) $\times10^{-12}$ & AGN & Y & 1 & 1 & 0 \\
8 & 4 & $10^h 13^m 05\fs43$ --$16^\circ 41' 21\farcs0$ & 6.7 & 0.016 ($\pm$0.008) & 7 ($\pm$3) $\times10^{-13}$ & X & Y & 0 & 0 & 0 \\
8 & 4 & $10^h 15^m 03\fs40$ --$16^\circ 52' 10\farcs4$ & 4.8 & 0.034 ($\pm$0.011) & 1.4 ($\pm$0.5) $\times10^{-12}$ & QSO & Y & 0 & 1 & 0 \\
8 & 4 & $10^h 15^m 56\fs13$ --$20^\circ 02' 28\farcs3$ & 5.6 & 0.17 ($\pm$0.06) & 7.1 ($\pm$2.6) $\times10^{-12}$ & Seyfert 1 & Y & 2 & 1 & 0 \\
8 & 2 & $10^h 18^m 47\fs53$ --$18^\circ 32' 39\farcs2$ & 5.1 & 0.05 ($\pm$0.03) & 2.0 ($\pm$1.5) $\times10^{-12}$ & Unknown & N & 0 & 0 & 4.3 \\
8 & 4 & $10^h 23^m 34\fs73$ --$19^\circ 32' 35\farcs6$ & 5.2 & 0.05 ($\pm$0.04) & 2.3 ($\pm$1.7) $\times10^{-12}$ & Star & Y & 2 & 1 & 1.3 \\
8 & 4 & $10^h 24^m 52\fs73$ --$19^\circ 56' 14\farcs2$ & 6.8 & 0.018 ($\pm$0.008) & 8 ($\pm$3) $\times10^{-13}$ & X & Y & 0 & 1 & 0 \\
8 & 4 & $10^h 50^m 57\fs19$ --$28^\circ 50' 00\farcs5$ & 5.1 & 0.018 ($\pm$0.008) & 8 ($\pm$4) $\times10^{-13}$ & Star & Y & 0 & 1 & 0 \\
8 & 4 & $11^h 18^m 15\fs20$ --$32^\circ 48' 12\farcs2$ & 6.5 & 0.020 ($\pm$0.008) & 9 ($\pm$3) $\times10^{-13}$ & Unknown & Y & 0 & 0 & 0 \\
8 & 4 & $11^h 19^m 45\fs86$ --$34^\circ 35' 51\farcs5$ & 6.8 & 0.019 ($\pm$0.009) & 8 ($\pm$4) $\times10^{-13}$ & X & Y & 1 & 1 & 0 \\
8 & 4 & $11^h 36^m 16\fs20$ --$38^\circ 02' 08\farcs2$ & 6.8 & 0.09 ($\pm$0.04) & 3.9 ($\pm$1.5) $\times10^{-12}$ & RSCVn & Y & 1 & 2 & 0 \\
8 & 4 & $11^h 39^m 01\fs66$ --$37^\circ 44' 17\farcs2$ & 4.2 & 0.98 ($\pm$0.19) & 4.2 ($\pm$0.8) $\times10^{-11}$ & Seyfert 1 & Y & 0 & 1 & 1.1 \\
13 & 4 & $11^h 42^m 47\fs84$ --$35^\circ 48' 57\farcs1$ & 7.1 & 0.049 ($\pm$0.012) & 2.1 ($\pm$0.5) $\times10^{-12}$ & Unknown & Y & 3 & 0 & 0 \\
14 & 4 & $11^h 47^m 56\fs31$ --$38^\circ 58' 13\farcs2$ & 6.4 & 0.09 ($\pm$0.06) & 4.1 ($\pm$2.4) $\times10^{-12}$ & Unknown & Y & 0 & 1 & 0 \\
8 & 4 & $12^h 10^m 04\fs18$ --$46^\circ 36' 24\farcs0$ & 4.3 & 0.113 ($\pm$0.020) & 4.9 ($\pm$0.8) $\times10^{-12}$ & Seyfert 2 & Y & 1 & 1 & 0 \\
8 & 4 & $12^h 11^m 15\fs32$ --$46^\circ 41' 31\farcs8$ & 7 & 0.022 ($\pm$0.009) & 9 ($\pm$4) $\times10^{-13}$ & Unknown & Y & 5 & 1 & 0 \\
8 & 3 & $12^h 23^m 18\fs14$ --$44^\circ 37' 31\farcs4$ & 6.4 & 0.031 ($\pm$0.012) & 1.3 ($\pm$0.5) $\times10^{-12}$ & Unknown & N & 0 & 1 & 1.1 \\
14 & 4 & $12^h 45^m 57\fs78$ --$12^\circ 51' 18\farcs0$ & 7.1 & 0.066 ($\pm$0.028) & 2.8 ($\pm$1.2) $\times10^{-12}$ & Galaxy & Y & 0 & 1 & 0 \\
14 & 4 & $12^h 49^m 10\fs81$ --$11^\circ 49' 24\farcs3$ & 5.9 & 0.08 ($\pm$0.03) & 3.4 ($\pm$1.3) $\times10^{-12}$ & Galaxy & Y & 4 & 1 & 0 \\
14 & 4 & $12^h 52^m 12\fs65$ --$13^\circ 24' 52\farcs6$ & 4.4 & 0.31 ($\pm$0.06) & 1.3 ($\pm$0.2) $\times10^{-11}$ & Radio (AGN; MQ) & Y & 4 & 1 & 0 \\
14 & 4 & $13^h 00^m 54\fs83$ --$21^\circ 34' 20\farcs6$ & 5.7 & 0.054 ($\pm$0.025) & 2.3 ($\pm$1.1) $\times10^{-12}$ & Unknown & Y & 0 & 1 & 0 \\
14 & 4 & $13^h 12^m 31\fs88$ --$21^\circ 56' 18\farcs9$ & 5.8 & 0.070 ($\pm$0.029) & 3.0 ($\pm$1.3) $\times10^{-12}$ & BLLac & Y & 0 & 1 & 0 \\
14 & 3 & $13^h 25^m 07\fs31$ --$32^\circ 38' 14\farcs4$ & 5.5 & 0.14 ($\pm$0.04) & 6.2 ($\pm$1.8) $\times10^{-12}$ & Radio & N & 0 & 1 & 0 \\
14 & 4 & $13^h 26^m 16\fs69$ --$29^\circ 05' 13\farcs0$ & 5.3 & 0.08 ($\pm$0.03) & 3.6 ($\pm$1.4) $\times10^{-12}$ & RotV* & Y & 3 & 1 & 0 \\
14 & 4 & $13^h 35^m 53\fs88$ --$34^\circ 17' 41\farcs2$ & 3.8 & 1.78 ($\pm$0.20) & 7.7 ($\pm$0.8) $\times10^{-11}$ & Seyfert 1 & Y & 8 & 1 & 0 \\
14 & 4 & $13^h 36^m 39\fs33$ --$33^\circ 57' 58\farcs0$ & 6.3 & 0.07 ($\pm$0.03) & 3.1 ($\pm$1.4) $\times10^{-12}$ & AGN & Y & 7 & 1 & 1.2 \\
14 & 4 & $13^h 38^m 23\fs70$ --$36^\circ 14' 01\farcs6$ & 4.7 & 0.17 ($\pm$0.05) & 7.3 ($\pm$2.0) $\times10^{-12}$ & Unknown & Y & 3 & 0 & 0 \\
14 & 4 & $14^h 10^m 34\fs05$ --$52^\circ 19' 06\farcs9$ & 6.1 & 0.11 ($\pm$0.03) & 4.8 ($\pm$1.5) $\times10^{-12}$ & GinGroup & Y & 4 & 1 & 0 \\
8 & 4 & $14^h 35^m 47\fs28$ --$52^\circ 40' 50\farcs9$ & 5.9 & 0.19 ($\pm$0.07) & 8.0 ($\pm$2.9) $\times10^{-12}$ & Star & Y & 1 & 1 & 0 \\
8 & 3 & $14^h 45^m 41\fs86$ --$49^\circ 22' 28\farcs6$ & 6.7 & 0.019 ($\pm$0.008) & 8 ($\pm$3) $\times10^{-13}$ & Unknown & N & 1 & 0 & 0 \\
8 & 4 & $15^h 10^m 05\fs74$ --$50^\circ 51' 33\farcs7$ & 5.9 & 0.09 ($\pm$0.05) & 4.0 ($\pm$2.1) $\times10^{-12}$ & Unknown & Y & 2 & 4 & 2 \\
8 & 3 & $15^h 23^m 17\fs80$ --$50^\circ 27' 03\farcs6$ & 6.2 & 0.049 ($\pm$0.011) & 2.1 ($\pm$0.5) $\times10^{-12}$ & Unknown & N & 0 & 3 & 0 \\
17 & 4 & $16^h 27^m 04\fs41$ +$14^\circ 21' 23\farcs3$ & 5.9 & 0.08 ($\pm$0.05) & 3.4 ($\pm$2.1) $\times10^{-12}$ & Seyfert 1 & Y & 2 & 1 & 0 \\
17 & 4 & $16^h 44^m 39\fs08$ --$01^\circ 51' 55\farcs5$ & 4.7 & 0.037 ($\pm$0.012) & 1.6 ($\pm$0.5) $\times10^{-12}$ & Unknown & Y & 1 & 0 & 0 \\
17 & 4 & $16^h 58^m 41\fs90$ --$03^\circ 14' 15\farcs2$ & 5.8 & 0.027 ($\pm$0.010) & 1.1 ($\pm$0.4) $\times10^{-12}$ & AGN Candidate & Y & 1 & 2 & 0 \\
17 & 3 & $18^h 22^m 20\fs87$ --$24^\circ 30' 04\farcs3$ & 6.9 & 0.022 ($\pm$0.009) & 10 ($\pm$4) $\times10^{-13}$ & Radio & N & 0 & 3 & 0 \\
17 & 3 & $18^h 28^m 06\fs37$ --$26^\circ 45' 24\farcs0$ & 6 & 0.025 ($\pm$0.010) & 1.1 ($\pm$0.4) $\times10^{-12}$ & ** & N & 0 & 1 & 0 \\
5 & 4 & $18^h 40^m 38\fs99$ --$77^\circ 09' 32\farcs7$ & 5.1 & 0.21 ($\pm$0.06) & 8.9 ($\pm$2.6) $\times10^{-12}$ & Radio(cm) & Y & 10 & 1 & 0 \\
17 & 3 & $18^h 41^m 36\fs88$ --$26^\circ 54' 18\farcs7$ & 6.5 & 0.013 ($\pm$0.005) & 5.5 ($\pm$2.0) $\times10^{-13}$ & Unknown & N & 0 & 2 & 0 \\
17 & 3 & $18^h 41^m 42\fs47$ --$31^\circ 11' 33\farcs8$ & 4.6 & 0.050 ($\pm$0.012) & 2.1 ($\pm$0.5) $\times10^{-12}$ & Unknown & N & 0 & 1 & 0 \\
17 & 3 & $18^h 41^m 47\fs10$ --$31^\circ 10' 05\farcs7$ & 5.5 & 0.029 ($\pm$0.010) & 1.2 ($\pm$0.4) $\times10^{-12}$ & Unknown & N & 0 & 1 & 0 \\
7 & 4 & $19^h 59^m 28\fs64$ +$40^\circ 44' 13\farcs0$ & 8.2 & 2.7 ($\pm$0.5) & 1.16 ($\pm$0.23) $\times10^{-10}$ & Seyfert 2 & Y & 4 & 2 & 0 \\
17 & 4 & $20^h 33^m 00\fs32$ --$34^\circ 40' 08\farcs9$ & 6 & 0.017 ($\pm$0.006) & 7.5 ($\pm$2.5) $\times10^{-13}$ & Unknown & Y & 0 & 1 & 0 \\
17 & 4 & $20^h 36^m 08\fs50$ --$36^\circ 07' 09\farcs5$ & 4.2 & 0.091 ($\pm$0.016) & 3.9 ($\pm$0.7) $\times10^{-12}$ & RotV* & Y & 2 & 1 & 0 \\
8 & 3 & $20^h 56^m 28\fs43$ +$30^\circ 45' 17\farcs6$ & 6.1 & 0.025 ($\pm$0.012) & 1.1 ($\pm$0.5) $\times10^{-12}$ & Unknown & N & 0 & 0 & 1.7 \\
8 & 4 & $20^h 58^m 12\fs30$ +$30^\circ 04' 36\farcs9$ & 4.3 & 0.100 ($\pm$0.017) & 4.3 ($\pm$0.7) $\times10^{-12}$ & Seyfert 1 & Y & 1 & 2 & 0 \\
8 & 4 & $20^h 58^m 52\fs39$ +$31^\circ 30' 12\farcs2$ & 5.8 & 0.017 ($\pm$0.006) & 7.5 ($\pm$2.7) $\times10^{-13}$ & Unknown & Y & 0 & 2 & 0 \\
8 & 3 & $21^h 11^m 42\fs05$ +$32^\circ 59' 25\farcs1$ & 7.2 & 0.019 ($\pm$0.008) & 8 ($\pm$4) $\times10^{-13}$ & Radio & N & 1 & 3 & 0 \\
8 & 3 & $21^h 11^m 55\fs13$ +$32^\circ 44' 46\farcs9$ & 6.4 & 0.031 ($\pm$0.011) & 1.3 ($\pm$0.5) $\times10^{-12}$ & Unknown & N & 0 & 1 & 1.3 \\
8 & 4 & $21^h 13^m 44\fs93$ +$35^\circ 31' 52\farcs4$ & 6.1 & 0.027 ($\pm$0.010) & 1.1 ($\pm$0.4) $\times10^{-12}$ & Unknown & Y & 0 & 1 & 0 \\
8 & 3 & $21^h 17^m 11\fs49$ +$36^\circ 04' 31\farcs9$ & 5.6 & 0.023 ($\pm$0.009) & 10 ($\pm$4) $\times10^{-13}$ & Star & N & 0 & 2 & 0 \\
8 & 4 & $21^h 21^m 01\fs16$ +$40^\circ 20' 36\farcs2$ & 4.5 & 0.08 ($\pm$0.04) & 3.3 ($\pm$1.5) $\times10^{-12}$ & SB* & Y & 1 & 1 & 0 \\
8 & 3 & $21^h 24^m 30\fs89$ +$40^\circ 15' 58\farcs4$ & 6.2 & 0.020 ($\pm$0.005) & 8.6 ($\pm$2.1) $\times10^{-13}$ & Unknown & N & 1 & 1 & 0 \\
8 & 4 & $21^h 27^m 14\fs46$ +$39^\circ 12' 34\farcs5$ & 6 & 0.016 ($\pm$0.007) & 6.8 ($\pm$2.9) $\times10^{-13}$ & Star & Y & 2 & 1 & 0 \\
8 & 3 & $21^h 33^m 14\fs37$ +$39^\circ 41' 07\farcs0$ & 5.5 & 0.06 ($\pm$0.04) & 2.8 ($\pm$1.7) $\times10^{-12}$ & Unknown & N & 0 & 1 & 1.6 \\
8 & 3 & $21^h 39^m 54\fs85$ +$44^\circ 45' 51\farcs1$ & 4.9 & 0.11 ($\pm$0.05) & 4.7 ($\pm$2.2) $\times10^{-12}$ & Unknown & N & 1 & 1 & 1.7 \\
8 & 4 & $21^h 42^m 43\fs28$ +$43^\circ 35' 12\farcs3$ & 4.2 & 2.7 ($\pm$0.4) & 1.17 ($\pm$0.18) $\times10^{-10}$ & DwarfNova & Y & 10 & 1 & 0 \\
8 & 4 & $21^h 53^m 15\fs93$ +$47^\circ 43' 48\farcs6$ & 5.2 & 0.023 ($\pm$0.010) & 10 ($\pm$4) $\times10^{-13}$ & Star & Y & 0 & 1 & 0 \\
8 & 4 & $22^h 04^m 56\fs63$ +$47^\circ 14' 08\farcs4$ & 5.5 & 0.15 ($\pm$0.06) & 6.6 ($\pm$2.4) $\times10^{-12}$ & RSCVn & Y & 0 & 1 & 0 \\
8 & 3 & $22^h 07^m 29\fs62$ +$49^\circ 31' 01\farcs1$ & 6.2 & 0.024 ($\pm$0.008) & 1.0 ($\pm$0.3) $\times10^{-12}$ & Unknown & N & 2 & 1 & 0 \\
8 & 4 & $22^h 07^m 47\fs03$ +$49^\circ 31' 44\farcs3$ & 5.3 & 0.021 ($\pm$0.008) & 9 ($\pm$3) $\times10^{-13}$ & Star & Y & 3 & 1 & 1 \\
8 & 3 & $22^h 08^m 13\fs89$ +$53^\circ 06' 51\farcs3$ & 5.9 & 0.016 ($\pm$0.007) & 7.1 ($\pm$3.0) $\times10^{-13}$ & Unknown & N & 0 & 2 & 0 \\
8 & 4 & $22^h 08^m 54\fs01$ +$50^\circ 50' 27\farcs4$ & 5.3 & 0.022 ($\pm$0.008) & 9 ($\pm$3) $\times10^{-13}$ & Star & Y & 1 & 1 & 0 \\
8 & 4 & $22^h 15^m 54\fs39$ +$52^\circ 18' 37\farcs6$ & 6.1 & 0.015 ($\pm$0.006) & 6.5 ($\pm$2.8) $\times10^{-13}$ & Unknown & Y & 0 & 2 & 0 \\
8 & 4 & $22^h 20^m 06\fs04$ +$53^\circ 00' 37\farcs6$ & 8.6 & 0.015 ($\pm$0.007) & 6.3 ($\pm$3.0) $\times10^{-13}$ & Unknown & Y & 1 & 6 & 0 \\
8 & 4 & $22^h 20^m 06\fs76$ +$49^\circ 30' 13\farcs3$ & 8.7 & 0.15 ($\pm$0.06) & 6.6 ($\pm$2.5) $\times10^{-12}$ & BYDra & Y & 0 & 1 & 0 \\
8 & 2 & $22^h 21^m 28\fs16$ +$50^\circ 32' 44\farcs2$ & 6.2 & 0.015 ($\pm$0.007) & 6 ($\pm$3) $\times10^{-13}$ & Unknown & N & 0 & 1 & 4.4 \\
8 & 3 & $22^h 28^m 29\fs95$ +$53^\circ 44' 10\farcs5$ & 5.4 & 0.034 ($\pm$0.010) & 1.4 ($\pm$0.4) $\times10^{-12}$ & Unknown & N & 0 & 1 & 5.2 \\
8 & 4 & $22^h 29^m 22\fs74$ +$53^\circ 49' 44\farcs5$ & 5.3 & 0.032 ($\pm$0.010) & 1.4 ($\pm$0.4) $\times10^{-12}$ & Unknown & Y & 1 & 0 & 0 \\
8 & 3 & $22^h 32^m 11\fs72$ +$54^\circ 08' 11\farcs2$ & 4.7 & 0.021 ($\pm$0.008) & 9 ($\pm$3) $\times10^{-13}$ & Unknown & N & 1 & 2 & 0 \\
8 & 4 & $22^h 32^m 37\fs82$ +$54^\circ 05' 29\farcs5$ & 5.6 & 0.067 ($\pm$0.014) & 2.9 ($\pm$0.6) $\times10^{-12}$ & Star & Y & 1 & 1 & 0 \\
8 & 4 & $22^h 35^m 40\fs00$ +$53^\circ 45' 31\farcs3$ & 6.2 & 0.05 ($\pm$0.04) & 2.3 ($\pm$1.8) $\times10^{-12}$ & AGN Candidate & Y & 1 & 1 & 1.2 \\
4 & 4 & $23^h 21^m 15\fs75$ --$26^\circ 58' 55\farcs8$ & 7.6 & 0.26 ($\pm$0.06) & 1.12 ($\pm$0.26) $\times10^{-11}$ & Unknown & Y & 1 & 0 & 0 \\
\enddata
\tablenotetext{}{The following columns are given:  \\
GW \# -- the number of the GW event (4 corresponds to the first entry in Table \ref{tab:obs_overview}, and 17 corresponds to last); \\
Rank -- as described in Section 3 (2 = interesting source, 3 = uncatalogued X-ray source, 4 = catalogued X-ray source); \\
RA/Dec (J2000); \\
Err -- 90\% positional uncertainty; \\
Peak Rate -- peak source XRT count rate; \\
Peak Flux -- (see Section 3 for details); \\
Simbad src type -- (see http://simbad.u-strasbg.fr/Pages/guide/chF.htx for details; ``AGN; MQ'' and ``AGN; S/W'' denote that there is a positionally coincident entry in the Million Quasars catalog or the QSO selection from SDSS and WISE; see \citealt{Flesch2015} and \citealt{Richards2015}, respectively); \\
Known X -- whether or not the source has been previously detected in X-rays; \\
Near gal.\ -- the number of known ``nearby'' galaxies (see Section 4.1 for details); \\
Near 2MASS -- the number of nearby / positionally coincident 2MASS sources; \\
Fading -- the statistical significance of fading behavior, if present.}
\end{deluxetable*}

%% file: swifto2.bbl
\begin{thebibliography}{ }

\bibitem[Abbott et al.(2017)]{Abbott2017} Abbott, B.~P., Abbott, R., Abbott, T.~D., et al.\ 2017, \apjl, 848, L12

\bibitem[Abbott et al.(2019)]{Abbott2019} Abbott, B.~P., Abbott, R., Abbott, T.~D., et al.\ 2019, \apj, 875, 161 

\bibitem[Acernese et al.(2015)]{VIRGO2015} Acernese, F., Agathos, M., Agatsuma, K., et al.\ 2015, Classical and Quantum Gravity, 32, 024001 

\bibitem[Barthelmy et al.(2005)]{Barthelmy2005} Barthelmy, S.~D., Barbier, L.~M., Cummings, J.~R., et al.\ 2005, \ssr, 120, 143 

\bibitem[Beniamini \& van der Horst(2017)]{Beniamini2017} Beniamini, P., \& van der Horst, A.~J.\ 2017, \mnras, 472, 3161 

\bibitem[Beniamini et al.(2019)]{Beniamini2019} Beniamini, P., Petropoulou, M., Barniol Duran, R., \& Giannios, D.\ 2019, \mnras, 483, 840 

\bibitem[Berger et al.(2003)]{Berger2003} Berger, E., Kulkarni, S.~R., \& Frail, D.~A.\ 2003, \apj, 590, 379 

\bibitem[Berger(2014)]{Berger2014} Berger, E.\ 2014, \araa, 52, 43 

\bibitem[Bilicki et al.(2014)]{Bilicki2014} Bilicki, M., Jarrett, T.~H., Peacock, J.~A., Cluver, M.~E., \& Steward, L.\ 2014, \apjs, 210, 9 

\bibitem[Bloom et al.(2001)]{Bloom2001} Bloom, J.~S., Frail, D.~A., \& Sari, R.\ 2001, \aj, 121, 2879 

\bibitem[Bulik et al.(1999)]{Bulik1999} Bulik, T., Belczy{\'n}ski, K., \& Zbijewski, W.\ 1999, \mnras, 309, 629 

\bibitem[Burrows et al.(2005)]{Burrows2005} Burrows, D.~N., Hill, J.~E., Nousek, J.~A., et al.\ 2005, \ssr, 120, 165 

\bibitem[D'Avanzo(2015)]{DAvanzo2015} D'Avanzo, P.\ 2015, Journal of High Energy Astrophysics, 7, 73 

\bibitem[Eichler et al.(1989)]{Eichler1989} Eichler, D., Livio, M., Piran, T., \& Schramm, D.~N.\ 1989, \nat, 340, 126 

\bibitem[Evans et al.(2009)]{Evans2009} Evans, P.~A., Beardmore, A.~P., Page, K.~L., et al.\ 2009, \mnras, 397, 1177 

\bibitem[Evans et al.(2014)]{Evans2014} Evans, P.~A., Osborne, J.~P., Beardmore, A.~P., et al.\ 2014, \apjs, 210, 8 

\bibitem[Evans et al.(2015)]{Evans2015} Evans, P.~A., Osborne, J.~P., Kennea, J.~A., et al.\ 2015, \mnras, 448, 2210 

\bibitem[Evans et al.(2016a)]{Evans2016a} Evans, P.~A., Osborne, J.~P., Kennea, J.~A., et al.\ 2016a, \mnras, 455, 1522

\bibitem[Evans et al.(2016b)]{Evans2016b} Evans, P.~A., Kennea, J.~A., Barthelmy, S.~D., et al.\ 2016b, \mnras, 460, L40 

\bibitem[Evans et al.(2016c)]{Evans2016c} Evans, P.~A., Kennea, J.~A., Palmer, D.~M., et al.\ 2016c, \mnras, 462, 1591 

\bibitem[Evans et al.(2017)]{Evans2017} Evans, P.~A., Cenko, S.~B., Kennea, J.~A., et al.\ 2017, Science, 358, 1565

\bibitem[Evans et al.(2019)]{Evans2019} Evans, P.~A., Kennea, J.~A., Palmer, D.~M., et al.\ 2019, \mnras, 484, 2362

\bibitem[Flesch(2015)]{Flesch2015} Flesch, E.~W.\ 2015, \pasa, 32, e010 

\bibitem[Fong et al.(2010)]{Fong2010} Fong, W., Berger, E., \& Fox, D.~B.\ 2010, \apj, 708, 9 

\bibitem[Fong et al.(2013)]{Fong2013} Fong, W., Berger, E., Chornock, R., et al.\ 2013, \apj, 769, 56

\bibitem[Fong et al.(2015)]{Fong2015} Fong, W., Berger, E., Margutti, R., \& Zauderer, B.~A.\ 2015, \apj, 815, 102 

\bibitem[Frail et al.(2001)]{Frail2001} Frail, D.~A., Kulkarni, S.~R., Sari, R., et al.\ 2001, \apjl, 562, L55 

\bibitem[Gehrels et al.(2004)]{Gehrels2004} Gehrels, N., Chincarini, G., Giommi, P., et al.\ 2004, \apj, 611, 1005 

\bibitem[Ghirlanda et al.(2019)]{Ghirlanda2019} Ghirlanda, G., Salafia, O.~S., Paragi, Z., et al.\ 2019, Science, 363, 968 

\bibitem[Haggard et al.(2017)]{Haggard2017} Haggard, D., Nynka, M., Ruan, J.~J., et al.\ 2017, \apjl, 848, L25 

\bibitem[HEASARC(2014)]{HEASARC2014} HEASARC 2014, ascl:1408.004

\bibitem[Kamble \& Kaplan(2013)]{Kamble2013} Kamble, A., \& Kaplan, D.~L.~A.\ 2013, International Journal of Modern Physics D, 22, 1341011 

\bibitem[Kanner et al.(2013)]{Kanner2013} Kanner, J., Baker, J., Blackburn, L., et al.\ 2013, \apj, 774, 63 

\bibitem[Li \& Paczy{\'n}ski(1998)]{Li1998} Li, L.-X., \& Paczy{\'n}ski, B.\ 1998, \apjl, 507, L59 

\bibitem[The LIGO Scientific Collaboration et al.(2015)]{LIGO2015} LIGO Scientific Collaboration, Aasi, J., Abbott, B.~P., et al.\ 2015, Classical and Quantum Gravity, 32, 074001 

\bibitem[The LIGO Scientific Collaboration et al.(2018)]{LVC2018} The LIGO Scientific Collaboration, the Virgo Collaboration, Abbott, B.~P., et al.\ 2018, arXiv:1811.12907

\bibitem[The LIGO Scientific Collaboration et al.(2019)]{LVC2019} The LIGO Scientific Collaboration, The Virgo Collaboration, \& authors, a.\ 2019, arXiv:1907.01443 

\bibitem[Liu et al.(2016)]{Liu2016} Liu, T., Romero, G.~E., Liu, M.-L., \& Li, A.\ 2016, \apj, 826, 82 

\bibitem[Loeb(2016)]{Loeb2016} Loeb, A.\ 2016, \apjl, 819, L21 

\bibitem[Margutti et al.(2017)]{Margutti2017} Margutti, R., Berger, E., Fong, W., et al.\ 2017, \apjl, 848, L20 

\bibitem[Margutti et al.(2018)]{Margutti2018} Margutti, R., Alexander, K.~D., Xie, X., et al.\ 2018, \apjl, 856, L18 

\bibitem[Metzger et al.(2010)]{Metzger2010} Metzger, B.~D., Mart{\'{\i}}nez-Pinedo, G., Darbha, S., et al.\ 2010, \mnras, 406, 2650 

\bibitem[Metzger \& Berger(2012)]{Metzger2012} Metzger, B.~D., \& Berger, E.\ 2012, \apj, 746, 48 

\bibitem[Middei et al.(2017)]{Middei2017} Middei, R., Vagnetti, F., Bianchi, S., et al.\ 2017, \aap, 599, A82 

\bibitem[Nakar et al.(2002)]{Nakar2002} Nakar, E., Piran, T., \& Granot, J.\ 2002, \apj, 579, 699 

\bibitem[Perna et al.(2016)]{Perna2016} Perna, R., Lazzati, D., \& Giacomazzo, B.\ 2016, \apjl, 821, L18 


\bibitem[Racusin et al.(2011)]{Racusin2011} Racusin, J.~L., Oates, S.~R., Schady, P., et al.\ 2011, \apj, 738, 138 

\bibitem[Richards et al.(2015)]{Richards2015} Richards, G.~T., Myers, A.~D., Peters, C.~M., et al.\ 2015, \apjs, 219, 39.

\bibitem[Roming et al.(2005)]{Roming2005} Roming, P.~W.~A., Kennedy, T.~E., Mason, K.~O., et al.\ 2005, \ssr, 120, 95 

\bibitem[Ryan et al.(2019)]{Ryan2019} Ryan, G., van Eerten, H., Piro, L., \& Troja, E.\ 2019, arXiv:1909.11691 

\bibitem[Saxton et al.(2008)]{Saxton2008} Saxton, R.~D., Read, A.~M., Esquej, P., et al.\ 2008, \aap, 480, 611 

\bibitem[Schutz(2011)]{Schutz2011} Schutz, B.~F.\ 2011, Classical and Quantum Gravity, 28, 125023 

\bibitem[Sister{\'o} \& Castore de Sister{\'o}(1974)]{Sistero1974} Sister{\'o}, R.~F., \& Castore de Sister{\'o}, M.~E.\ 1974, \aj, 79, 391 

\bibitem[Shoemaker \& LIGO Scientific Collaboration(2019)]{Shoemaker2019} Shoemaker, D., \& LIGO Scientific Collaboration 2019, \baas, 51, 452 

\bibitem[Sun et al.(2017)]{Sun2017} Sun, H., Zhang, B., \& Gao, H.\ 2017, \apj, 835, 7 

\bibitem[Troja et al.(2017)]{Troja2017} Troja, E., Piro, L., van Eerten, H., et al.\ 2017, \nat, 551, 71 

\bibitem[Troja et al.(2018)]{Troja2018} Troja, E., Piro, L., Ryan, G., et al.\ 2018, \mnras, 478, L18 

\bibitem[Tunnicliffe et al.(2014)]{Tunnicliffe2014} Tunnicliffe, R.~L., Levan, A.~J., Tanvir, N.~R., et al.\ 2014, \mnras, 437, 1495 

\bibitem[van Eerten et al.(2010)]{vanEerten2010} van Eerten, H., Zhang, W., \& MacFadyen, A.\ 2010, \apj, 722, 235 

\bibitem[van Eerten(2018)]{vanEerten2018} van Eerten, H.\ 2018, International Journal of Modern Physics D, 27, 1842002-314 

\bibitem[Voges et al.(1999)]{Voges1999} Voges, W., Aschenbach, B., Boller, T., et al.\ 1999, \aap, 349, 389 

\bibitem[White et al.(2011)]{White2011} White, D.~J., Daw, E.~J., \& Dhillon, V.~S.\ 2011, Classical and Quantum Gravity, 28, 085016 

\bibitem[Yamazaki et al.(2016)]{Yamazaki2016} Yamazaki, R., Asano, K., \& Ohira, Y.\ 2016, Progress of Theoretical and Experimental Physics, 2016, 051E01 

\bibitem[Zhang(2016)]{Zhang2016} Zhang, B.\ 2016, \apjl, 827, L31 



\end{thebibliography}
